\documentclass[twocolumn]{svjour3}          
\usepackage[utf8]{inputenc}
\usepackage{xcolor} 
\usepackage{amsmath,amsfonts,amssymb}
\usepackage{hyperref} 
\usepackage{subcaption} 
\usepackage{url}
\usepackage{bm}
\usepackage{placeins} 
\usepackage{stfloats}
\usepackage{mwe}
\usepackage{array,multirow} 
\usepackage{mathrsfs}  
\usepackage{gensymb}   
\usepackage{natbib}

\makeatletter
\let\cl@chapter\undefined
\makeatletter
\usepackage[noabbrev]{cleveref}
    \Crefname{equation}{Eq.}{Eqs.}
\Crefname{figure}{Fig.}{Figs.}
\Crefname{tabular}{Tab.}{Tabs.}

\renewcommand{\cref}{\Cref}

 \usepackage[normalem]{ulem}
 \newcommand{\ADD}[1]{#1}
 \newcommand{\RM}[1]{}

 \newcommand{\RMmath}[1]{}

\title{\ADD{Simultaneous} shape and topology optimization of wings}
\author{Lukas C. Høghøj$^{a*}$\and Cian Conlan-Smith$^b$\and Ole Sigmund$^a$\and Casper Schousboe Andreasen$^a$}

\institute{$^a$Department of Civil \& Mechanical Engineering, Section for Solid Mechanics, Technical University of Denmark, 2800 Kgs. Lyngby, Denmark\and $^b$ Airbus Operations GmbH, Flight Physics Capabilities, 28199 Bremen, Germany
\and 
$^*$Corresponding author, \email{luch@dtu.dk}, Tel.: +4545254262 }
\date{Received: date / Accepted: date}

\begin{document}

\maketitle

\noindent Please cite the following version: \url{https://doi.org/10.1007/s00158-023-03569-x}\\

\begin{abstract}
This paper presents a method for simultaneous optimization of the outer shape and internal topology of aircraft wings, with the objective of minimizing drag subject to lift and compliance constraints for multiple load cases. The physics are evaluated by the means of a source-doublet panel method for the aerodynamic response and linear elastic finite elements for the structural response, which are one-way coupled. At each design iteration, a mapping procedure is applied to map the current wing shape and corresponding pressure loads to the unfitted finite element mesh covering the design domain. Wings of small fixed-wing airplanes both with and without a stiffening strut are optimized. The resulting wings show internal topologies with struts and wall-truss combinations, depending on the design freedom of the shape optimization. The lift distributions of the optimized wings show patterns \ADD{like} the ones obtained when performing optimization of wing shapes with constraints on the bending moment at the root. 
\end{abstract}

\section{Introduction}
When designing aircraft wings, the structural and aerodynamic performance of the design \ADD{should} be considered simultaneously. High aspect ratios reduce the induced drag on the wing. However, they are inefficient from a structural point of view, as the distributed pressure loads create large bending moments. Structural stiffness is increased by adding material inside the wing which increases the mass. Naturally, this raises the question: "Where should the material be placed to get the most out of it?".

Topology optimization is a computational method to determine material layout in a design domain subject to some physics and boundary conditions. Since the seminal paper by \cite{Bendsoee1988}, the method has been applied to a wide range of physics and cases. The method is known for its large design freedom with no specific initial design required. Efforts have been placed on the topology optimization of the internal structure of wings. \ADD{\cite{Stanford2015}} investigated the topology of ribs and spars under \ADD{aeroelastic loads of a fixed outer wing shape.} \cite{Dunning2015} optimized the internal structure of a wing subject to aerodynamic loads, that depends on the deflection. Finally, \cite{Aage2017}, used over a billion elements to optimize the detailed internal structure of the NASA Common Research Model (CRM) wing. This showcased the need for a fine resolution to capture intricate and realistic details in the obtained structure.

The first application of numerical optimization to aeroelasticity was performed by \cite{haftka1977a}, where \ADD{trade-offs} between weight and drag were achieved in the optimized designs. Since then, \ADD{several} studies have been performed on the optimization of aeroelastic structures. Among others, \cite{Grossman1990} used lifting-line and beam theory to optimize sailplane wings. \cite{Maute2004} used the finite volume method for Euler flow with an Arbitrary Lagrangian-Eulerian formulation to optimize the internal structure of wings. \cite{Barcelos2008} used Finite Volumes for the turbulent flow and geometrically non-linear Finite Elements for the structure to optimize the wing back sweep and twist as well as the thickness of the internal stiffeners. \cite{Kennedy2014} used panel methods to optimize wing shape and internal stiffener layouts. \ADD{\cite{Stanford2021} optimized the wing shape, size, and layout parameters of a wing box consisting of ribs and spars.} \cite{Wang2020b,Wang2020a} optimized the outer shape and topology of beam cross sections of wind turbine blades and compared approaches optimizing the outer shape and inner beam simultaneously and sequentially. The coupled optimization turned out to give the best results. \cite{Conlan-Smith2021} optimized the aerostructural problem using panel methods and beams for the internal structure modeling, where the difference between straight and curved spars was investigated.  \cite{James2014} used panel methods to optimize the wing shape \ADD{by the twist} and the internal topology \ADD{of the wing box}.  \cite{Gomes2020} solved the RANS equations with SST to optimize the external shape and internal topology of airfoils. The same authors also applied similar methods to entire wings using the Spalart-Allmaras turbulence model where the internal topology is optimized with a soft and stiff material, while no void is allowed (\cite{Gomes2022}).

Panel methods are computational tools used to evaluate the performance of preliminary aircraft design. The methods utilize potential flow theory to obtain velocity and pressure fields around a given geometry. The method only requires a surface mesh of the analyzed geometry - making the method attractive for optimization settings due to its fast turnaround time and no need for large fluid domain meshes.

\ADD{Until now, no coupled shape and topology optimization has been performed on wings, where the internal topology is free to evolve into a structure of material and void. The present work optimizes the} outer aerodynamic shape and \ADD{full} internal structural topology of aircraft wings simultaneously. The wing skin is parameterized by NACA airfoil sections and the aerodynamic response of the wing is evaluated using source-doublet panel methods due to their fast turnaround time (\cite{Conlan-Smith2020}). The interior of the wing is modeled using linear elastic hexahedral finite elements. The pressure loads from the aerodynamic analysis are mapped from the surface mesh of the wing to the corresponding integration points in an unfitted finite element mesh, using a scheme similar to the one outlined by \cite{Lee2012}. The structural design is represented by a density variable in each element. \ADD{The wing skin is mapped on the structural mesh and set to a fixed thickness. In the interior of the wing, the density variables determine if a given element should be solid or void. Unlike in earlier works, the entirety of the wing structure is included in the optimization process, allowing for features to appear inside all of the wing and not limited to a wing box. Furthermore, the topology optimization representation contains an order of magnitude more elements than the current state of the art (\cite{Gomes2022}).} To limit the complexity and computational burden, the response of the wing is considered as one-way coupled, letting the aerodynamic performance be unaffected by the wing deformation, which is a reasonable approximation if the deflections of the wing are small (\cite{Conlan-Smith2022}). The goal of the optimization is to minimize the induced drag while having sufficient lift to carry the payload and wing structure as well as satisfying the compliance constraints under both cruise and takeoff conditions. Besides forcing the wing to \ADD{store a low amount of elastic energy}, the compliance constraints also \ADD{limit} the deformations of the final wing, ensuring that the one-way coupling is a reasonable approximation. \ADD{The presented work does not consider buckling constraints, which are highly relevant for the structural integrity of a wing, especially in the skin. However, at present, buckling constrained topology optimization is not yet at a stage where it can be applied to large-scale problems, such as the ones considered here. The largest three-dimensional topology optimization problems with buckling constraints were run with 2 million degrees of freedom in the work by \cite{Ferrari2020a}.} The optimization is performed on a fine structural mesh, with $27\cdot 10^6$ design variables, representing solid and void elements.

The remainder of this paper is structured as follows; \cref{sec:param} presents the parametrizations of the problem; \cref{sec:phys} presents the physics utilized in the model of the study; \cref{sec:opt} outlines the optimization formulation; \cref{sec:res} results; and \cref{sec:conc} provides a discussion and conclusions.

\section{Parametrization}
\label{sec:param}
\begin{figure*}[htb]
    \centering
    \begin{subfigure}[c]{0.35\linewidth}
    \includegraphics[width=\linewidth]{inside_dots.png}
    \end{subfigure}~
    \begin{subfigure}[c]{0.125\linewidth}
    \includegraphics[width=\linewidth]{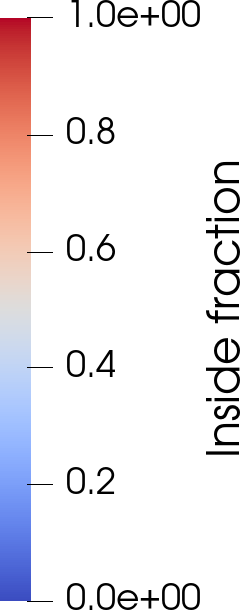}
    \end{subfigure}~
    \begin{subfigure}[c]{0.5\linewidth}
    \includegraphics[width=\linewidth]{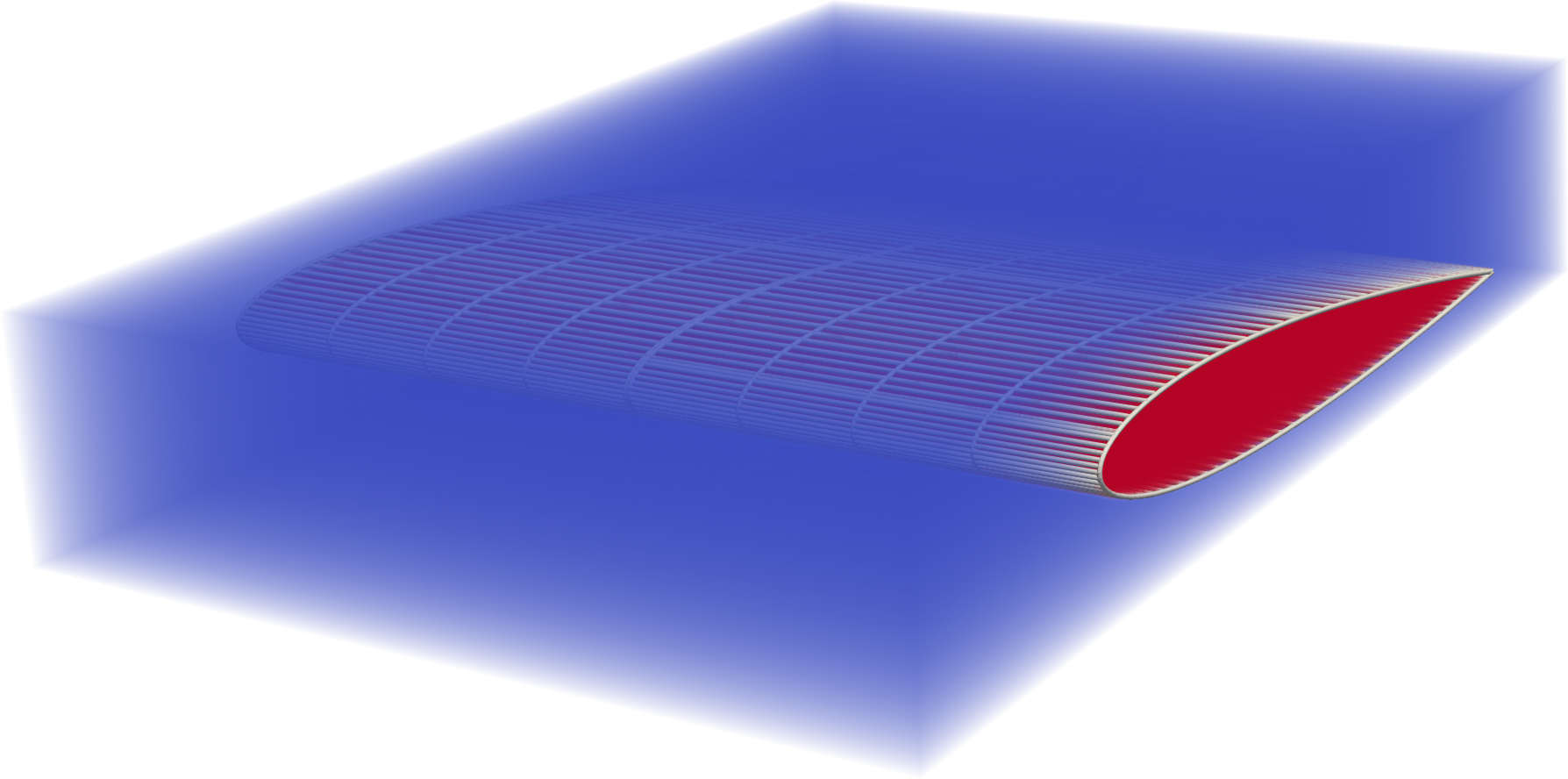}
    \end{subfigure}~
    \caption{Illustration in 2D (left) and 3D (right) of the two parametrizations, the white line represents the surface mesh of the wing, which is superposed on the fixed finite element mesh. The \ADD{colors} of the elements indicate what fraction of the element is inside the wing.}
    \label{fig:param}
\end{figure*}

Due to the difference in the nature of the modeling and optimization problems for the external shape of the wing and the internal structure, different parametrizations are used. \cref{fig:param} shows how the wing shape is modeled using a surface panel mesh, in the 2D view represented by a piecewise linear white curve (representing cuts along panels), and the internal structure is represented and modeled using an unfitted mesh. This choice is made to avoid remeshing operations at every design iteration; which would also require remapping the current design to a new mesh. Such operations would result in a high computational burden \ADD{and increased complexity of the presented method}. The unfitted mesh must surround the wing shape at every iteration step. The geometry of the unfitted mesh could be the design space of the wing shape. Computational efforts can however be saved if intuition or experience from past optimizations is used to make tighter meshes. If the wing is not fully included in the finite element mesh after a design update, the optimization process needs to be aborted and potentially reinitialized with a larger or repositioned mesh.

\subsection{Wing shape}
\label{sec:pandisc}
\begin{figure}[htb]
\centering
\includegraphics[width=\linewidth]{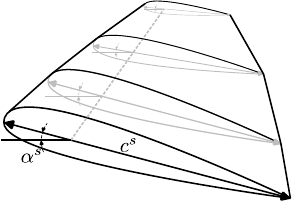}
\caption{Illustration of the optimized twist and chord length at each wing section.}
\label{fig:wing_para}
\end{figure}

The wing shape parametrization introduced in \cite{Conlan-Smith2020,Conlan-Smith2021} is used. The wing is decomposed in spanwise sections where each airfoil section is parameterized by NACA 4-digit airfoils (\cite{Abbot1959}). The local chord length, $c$, and twist, $\alpha$, are optimized, as shown in \cref{fig:wing_para}. The airfoil sections are linearly distributed along the span. The design variables of a section $d^s_i$, where $i$ indicates the shape parameter and $s$ the section, are all normalized by the gap between their minimum and maximum values, denoted by the subscripts $_\mathit{min}$ and $_\mathit{max}$, respectively, such that they all are between 0 and 1:
\begin{equation}
\label{eq:shapeinterp}
\begin{split}
&\alpha^s=\alpha_\mathit{min}+d^s_1\left(\alpha_\mathit{max}-\alpha_\mathit{min}\right)\\
&c^s=c_\mathit{min}+d^s_2\left(c_\mathit{max}-c_\mathit{min}\right)
\end{split}
\end{equation}

The shape design variables are regularized along the span using a filter of radius $r_a$, similar to a \ADD{one-dimensional} density filter from topology optimization (\cite{Bourdin2001,Bruns2001}). This is done to circumvent designs with heavy oscillations due to numerical artifacts of the panel method, as discussed by \cite{Conlan-Smith2020}.

Only the local twist and chord lengths are optimized in the present study. One could easily also include the relative thickness, maximum camber, and the position of the maximum camber, however, due to the used inviscid theory, this may lead to multiple designs that can achieve the same aerodynamic properties (\cite{Conlan-Smith2022}), and thus ill-posed optimization problems.

\subsection{Structural parametrization}
The wing is represented such that the internal topology can be optimized, yet, the skin should be solid, with a given thickness to guarantee that the aerodynamic loads are transferred to solid elements. If aerodynamic forces are transferred to void elements, the local compliance will become high\ADD{,} and the load transfer be \ADD{non-physical}. The elements outside the wing should be identified and only provide negligible stiffness to the structural model.

The parameterization of the wing structure is divided into three steps. First, the wing shape is mapped to the unfitted finite element mesh and the skin of the wing is fixed to be solid. The structure inside the wing is then parameterized using the structural design variable. These two parts are then agglomerated. An overview of the different steps in the process used to differentiate between skin and internal elements, as well as the combination of the design variables is illustrated in \cref{fig:var_aggreg}. Each of the parametrization steps is outlined in one of the following three subsections.

\subsubsection{Wing skin mapping}
The current wing shape is superimposed on the finite element mesh, which is not fitted to the current wing shape. An identification field, $\xi$, represents the element volume fraction inside of the wing shape. An erosion-dilation technique is applied on $\xi$, to obtain a fixed skin thickness on the wing, as was done in \cite{Hoghoj2020}, inspired by \cite{Luo2019}. 

The $\xi$ field is first filtered with a filter radius $r_e$, using the modified Helmholtz PDE filter by \cite{Lazarov2011}:
\begin{equation}
    \label{eq:filter}
    -\frac{r_e^2}{12}\nabla^2\tilde{\xi}+\tilde{\xi}=\xi
\end{equation}

The eroded and dilated fields are obtained by projecting the filtered field $\tilde{\xi}$ using a smooth Heaviside projection \cite{Wang2011}:
\begin{equation}
\label{eq:heavi}
    \hat{\xi}(\tilde{\xi},\beta_\mathit{skin},\eta) = \frac{\tanh(\beta_\mathit{skin}\eta)+\tanh(\beta_\mathit{skin}(\tilde{\xi}-\eta))}{\tanh(\beta_\mathit{skin}\eta)+\tanh(\beta_\mathit{skin}(1-\eta))}
\end{equation}
where $\beta_\mathit{skin}$ is the projection sharpness and $\eta$ the threshold value. A high and a low threshold value is used for the eroded- and dilated fields, respectively, $\eta=0.5\pm\Delta\eta$.

\begin{figure*}[htb]
    \centering
    \begin{subfigure}[t]{0.31\textwidth}
    \includegraphics[width=\linewidth]{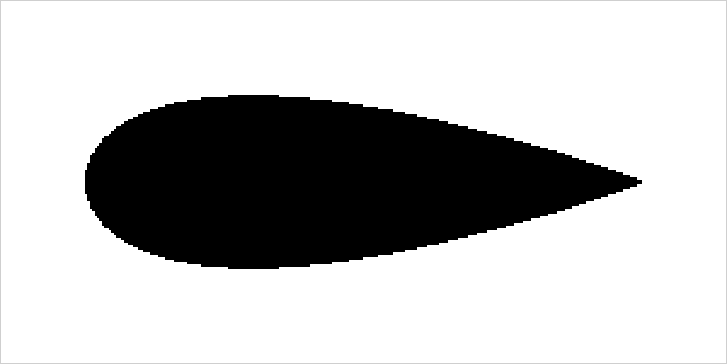}
    \caption{$\xi$}
    \label{fig:erodilinit}
    \end{subfigure}
    \begin{subfigure}[t]{0.31\textwidth}
    \includegraphics[width=\linewidth]{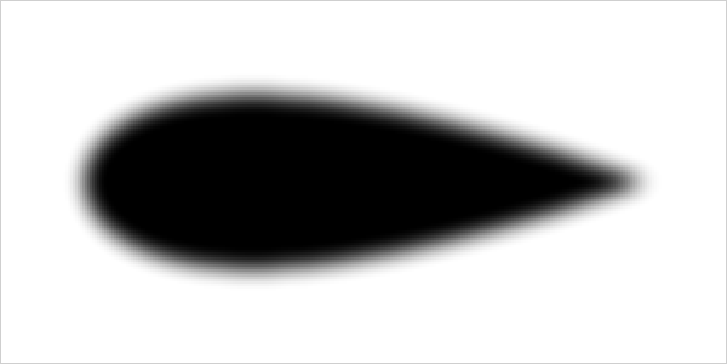}
    \caption{$\tilde{\xi}$}
    \end{subfigure}
    \begin{subfigure}[t]{0.31\textwidth}
    \includegraphics[width=\linewidth]{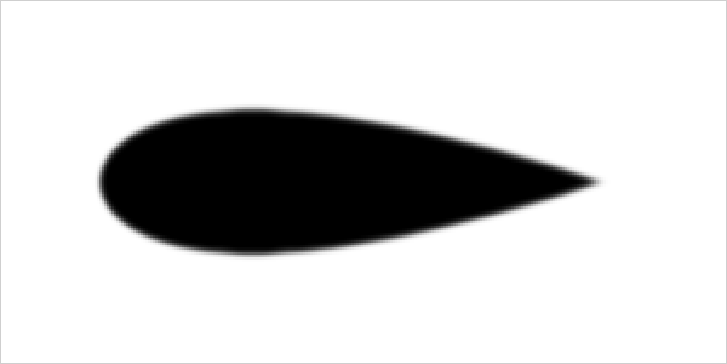}
    \caption{$\underline{\xi}$}
    \label{fig:erodilero}
    \end{subfigure}
    \\
    \begin{subfigure}[t]{0.31\textwidth}
    \includegraphics[width=\linewidth]{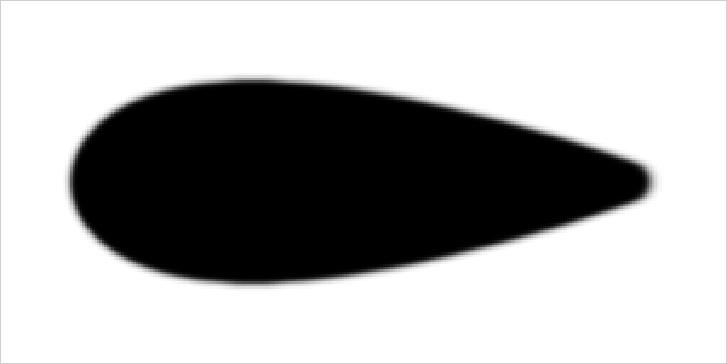}
    \caption{$\overline{\xi}$}
    \label{fig:erodildil}
    \end{subfigure}
    \begin{subfigure}[t]{0.31\textwidth}
    \includegraphics[width=\linewidth]{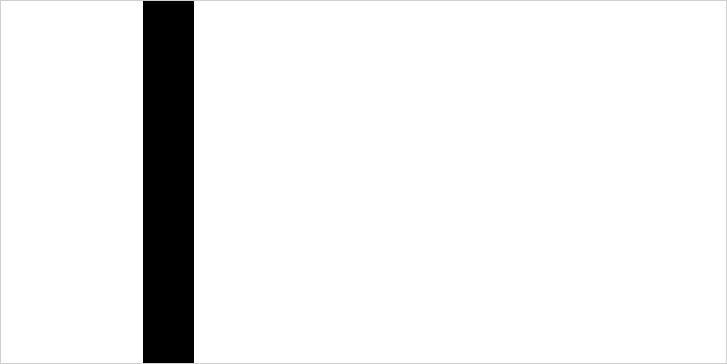}
    \caption{$\gamma$}
    \end{subfigure}
    \begin{subfigure}[t]{0.31\textwidth}
    \includegraphics[width=\linewidth]{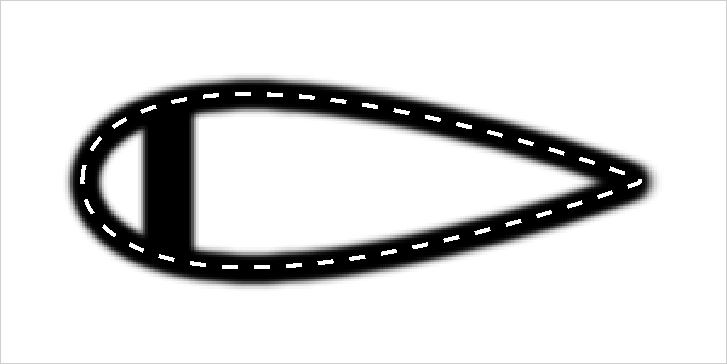}
    \caption{$\rho$}
    \end{subfigure}\\
    \begin{subfigure}[t]{0.3\textwidth}
    \includegraphics[width=\linewidth]{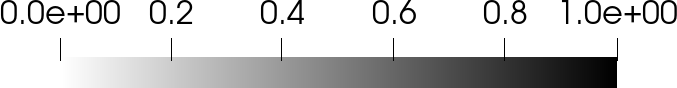}
    \end{subfigure}
    \caption{The wing identification field, $\xi$ (a), is filtered, $\tilde{\xi}$ (b), which is thresholded to the eroded $\underline{\xi}$ (c), and dilated $\overline{\xi}$ (d). Combined with the structural design variable, $\gamma$ (e), the physical design field, $\rho$ (f) is obtained using $\rho_\mathit{skin}=1$.}
    \label{fig:var_aggreg}
\end{figure*}

The eroded and dilated fields, $\underline{\xi}$ and $\overline{\xi}$, respectively, of the field $\xi$, seen in \cref{fig:erodilinit}, are shown in \cref{fig:erodilero,fig:erodildil}, respectively.  The wall is then identified by subtracting the eroded field from the dilated one, $\overline{\xi}-\underline{\xi}$.

A fixed wing skin thickness, $w_e$, can be achieved when setting the filter radius, $r_e$, and the threshold value gap $\Delta\eta$ for the smooth Heaviside functions according to \cite{Luo2019}:
\begin{align}
        r_e &\approx 0.75w_e & \Delta\eta&=0.45
\end{align}

It is noted that the wing skin should be at least two elements thick, such that $w_e$ is twice the average element \ADD{side length}, to ensure that only solid elements are loaded by the load transfer scheme discussed in \cref{sec:loadtrans}. 

\ADD{As seen in \cref{fig:var_aggreg}, the erosion-dilation process for the modeling of the wing skin places half of the skin material outside the wing shape from the shape parametrization discussed in \cref{sec:pandisc}. The outside of the skin could be aligned with the parametrization by using $\eta=0.95$ and $\eta=0.5$ for the eroded and dilated fields, respectively. The relation between skin thickness and filter radius would then be $r_e\approx1.5w_e$. However, only using the erosion step could be problematic, as this could result in loads being transferred to degrees of freedom only belonging to void elements.}

\subsubsection{Internal structure}
\label{sec:internal}

In addition to the identification functions, an element-wise structural design variable $\gamma$ is used to determine the state of the corresponding element. A structural design variable exists in each element of the finite element mesh, both the ones inside and outside of the current wing geometry. The structural design variable is filtered using the PDE filter \eqref{eq:filter} with filter radius $r_s$, leading to the filtered variable $\tilde{\gamma}$. The filtered variable is subsequently thresholded with the smooth Heaviside projection \eqref{eq:heavi} using high-, intermediate- and low threshold values, leading to eroded, nominal, and dilated fields, $\gamma_e$, $\gamma_n$ and $\gamma_d$, respectively. The eroded field is used for the computation of the structural response of the wing, while the dilated field is used to compute the mass of the structure. This approach ensures a crisp design field as the projection sharpness $\beta$ is ramped up, with a length scale imposed on it (\cite{Sigmund2009,Wang2011}). 

\subsubsection{Field agglomeration}
 The final physical design variable, $\rho$ is then obtained by combining the eroded- and dilated indicator variables, $\underline{\xi}$ and $\overline{\xi}$, respectively. The relative density of the wing skin $\rho_\mathit{skin}\in\left[0;\,1\right]$ and any state of the internal structural variable (eroded, nominal, or dilated) is obtained as:
\begin{equation}
\begin{split}
    \rho_e &= \rho_\mathit{skin}\left(\overline{\xi} - \underline{\xi}\right) + \underline{\xi}\gamma_e\\
    \rho_n &= \rho_\mathit{skin}\left(\overline{\xi} - \underline{\xi}\right) + \underline{\xi}\gamma_n\\
    \rho_d &= \rho_\mathit{skin}\left(\overline{\xi} - \underline{\xi}\right) + \underline{\xi}\gamma_d
\end{split}
\end{equation}

The stiffness $E$ is interpolated linearly from the eroded physical field $\rho_e$ using its upper and lower bounds, $E_\mathit{max}$ and $E_\mathit{min}$, respectively. The mass of the structure is interpolated linearly from the dilated physical field, $\rho_d$ and the material mass density $W_\mathit{structure}$:

\begin{equation}
\label{eq:simp}
\begin{split}
    E &= E_\mathit{min} + \rho_e\left(E_\mathit{max}-E_\mathit{min}\right)\\
    M &= W_\mathit{structure}\rho_d
\end{split}
\end{equation}

Remark that the usual SIMP penalization is not used here since the applied robust formulation naturally leads to black-and-white designs. Hence the penalization of intermediate design values is not necessary.
\section{Physical response} 
\label{sec:phys}
The \ADD{one-way} coupling allows for a sequential evaluation of the physics - first, the aerodynamic response is modeled using of the panel method; then the structural response is evaluated using linear elastic finite elements.
\subsection{Aerodynamic model}
\label{sec:panel}

\begin{figure*}[htb]
    \centering
    \includegraphics[width=\textwidth]{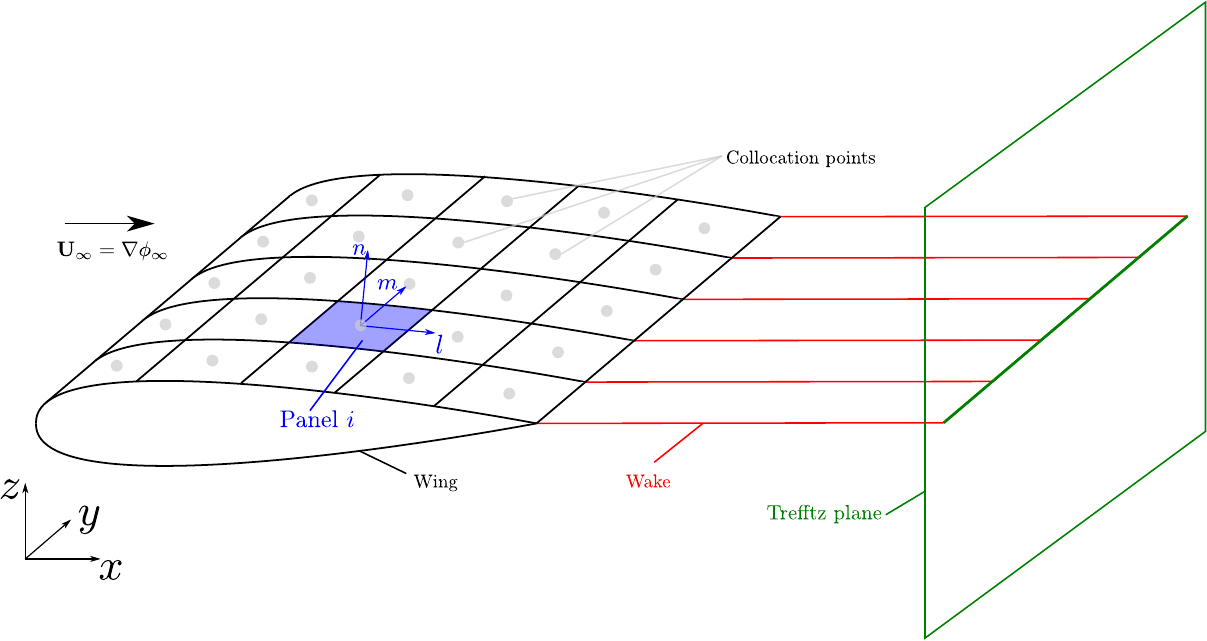}
    \caption{Sketch of the wing discretization with a surface mesh and the local panel coordinate system.}
    \label{fig:panel}
\end{figure*}

The aerodynamic response of the wing is modeled using a source-doublet panel method, as was done by \cite{Conlan-Smith2020}. The wing is represented by panels discretizing the space between the airfoil sections from the parametrization discussed in \cref{sec:pandisc}, as shown in the sketch in \cref{fig:panel}. Dirichlet boundary conditions are enforced such that the potential function inside the wing is constant and that there is no flow through the wing skin. The dense aerodynamic influence coefficient matrices A and B are calculated as outlined in \cite{katz2001a}. The doublet strengths, $\bm{\mu}$, are computed by solving:
\begin{equation}
\label{eq:aerostate}
    \mathbf{r}_a=\mathbf{A}\bm{\mu}+\mathbf{B}\bm{\sigma}=\mathbf{0}
\end{equation}
where the source strengths $\bm{\sigma}$ are obtained from the free stream velocity $\mathbf{U}_\infty$ and the normal of panel $i$, $\mathbf{n}_i$:
\begin{equation}
    \sigma_i=\mathbf{U}_\infty\cdot \mathbf{n}_i
\end{equation}

A fixed freestream wake as indicated in \cref{fig:panel} is also considered in the modeling of the aerodynamic response to ensure that the Kutta condition, stating that the flow must separate from the wing at the trailing edge with a finite velocity, is met.

The potential function $\phi$ can then be used to recover the local velocities, by computing its gradient, in the local coordinates $\left(l,\;m,\;n\right)$, of the potential function:
\begin{equation}
    \mathbf{U}_i = \left(\frac{\partial\phi}{\partial l},\,\frac{\partial\phi}{\partial m},\,\frac{\partial\phi}{\partial n}\right)^\intercal_i
\end{equation}
The panel pressure coefficients are recovered from the panel velocities, using the Bernoulli equation: 
\begin{equation}
    C_{P,\,i}=1-\frac{\left|\left|\mathbf{U}_i\right|\right|^2}{\left|\left|\mathbf{U}_\infty\right|\right|^2}
\end{equation}

 Drag and lift are computed through far-field computations on the Trefftz plane, where they are obtained by line integrals over the wake (\cite{Drela2014}): 
\begin{align}
    L &= \rho_\infty V_\infty\int_{S_w}\Delta\phi dy \approx \rho_\infty V_\infty\sum_{i=1}^{N_w}\mu_is_i\cos(\theta_i)\\
    D &= -\frac{1}{2}\rho_\infty\int_{S_w}\Delta\phi\frac{\partial\phi}{\partial n} dS_w \approx -\frac{1}{2}\rho_\infty \sum_{i=1}^{N_w}\mu_is_i\mathbf{u}_i\cdot\mathbf{n}_i
\end{align}
which ensures a robust computation of the forces. Note that only the induced drag is computed, as accurate methods for viscous drag computation are not easily introduced for potential flow models.
\subsection{Structural model}
The structural analysis is done using a linear elastic Finite Element model. Hexahedral elements are implemented in the unstructured topology optimization framework from \cite{Traff2021}. The residual to the finite element model is given as:
\begin{equation}
\label{eq:femRes}
    \mathbf{r}_s=\mathbf{K}\mathbf{u}-\mathbf{p}=\mathbf{0}
\end{equation}
 where $\mathbf{K}$ is the global stiffness matrix, $\mathbf{u}$ the displacement vector and $\mathbf{p}$ the load vector. The global stiffness matrix $\mathbf{K}$ is obtained after assembly of the element matrices $\mathbf{k}_e(\rho_e)$ which depend on the interpolation of the element's Young's modulus as outlined in \eqref{eq:simp}, using the assembly operator $\text{\Large $\mathbf{A}$}$, that adds up the corresponding degrees of freedom in the global matrix:
 \begin{equation}
     \mathbf{K}=\text{\Large $\mathbf{A}$}_{e=1}^{n_\mathit{el}}\mathbf{k_e}(\rho_e)
 \end{equation}
 
\subsection{Load transfer scheme}
\label{sec:loadtrans}
\begin{figure}[htb]
    \centering
    \includegraphics[width=\linewidth]{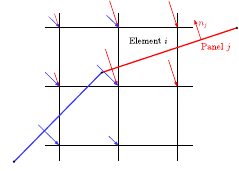}
    \caption{Sketch of the mapping process of pressure forces from the wing panel mesh to the unfitted finite element mesh. Note that forces are mapped to both nodes inside and outside of the wing.}
    \label{fig:presmap}
\end{figure}

The pressure loads obtained in the panel method are transferred to the nodes belonging to the finite elements, which are intersected by the wing geometry. The process is visualized in \cref{fig:presmap}. The area of the panel $j$ inside the finite element $i$ is computed. The pressure contributions of the wing panel $j$ on the finite element $i$ can then be projected to the corresponding finite element nodes using the outward pointing normal of the wing panel $j$, $\mathbf{n}_j$, as shown in \cref{fig:panel}\ADD{, and the area of the panel inside the element $i$, $A_{i\cap j}$. The force of the panel acting on the element is then projected to the nodes of the element by multiplying the resulting force with the shape functions of the finite element, evaluated at the center of the intersection of the panel and the element, $\mathbf{N}_{i\cap j}$:}
\begin{equation}
\label{eq:presmap}
    p_i = -\sum_{j=1}^{N_\mathit{panels}} p_j\mathbf{n}_j A_{i\cap j}\mathbf{N}_{i\cap j}
\end{equation}
This results in loads being placed on all nodes connected to an element that is intersected by the wing geometry, as seen in the sketch from \cref{fig:presmap}. Hence, a wing skin thickness of at least two elements should be chosen, to ensure that loads are transferred to solid elements. It should further be noted that gravity is not considered in the finite element model. This means that the design does not benefit from load alleviation.

Note that the applied load transfer scheme ensures consistency between the two employed models. This results from the loads transferred to the individual finite elements of the mesh being partial integrations of the pressure forces on the panels. The sum of all contributions from a given panel is hence equal to the pressure forces integrated over the panel area.

\section{Optimization formulation}
\label{sec:opt}
The goal of the coupled aerodynamic and structural response of the optimized wing is to minimize the drag under cruise conditions, $D_\mathit{cruise}$. A constraint is enforced such that the cruise lift, $L_\mathit{cruise}$, is greater than the weight of the wing, $W=\int_\Omega M(\rho_d)dV$, and the payload, $P$. Two structural constraints are enforced on the compliances of the wing at cruise and takeoff conditions, $C_\mathit{cruise}$ and $C_\mathit{takeoff}$, respectively, with the upper bounds $\overline{C}_\mathit{cruise}$ and $\overline{C}_\mathit{takeoff}$\RM{ to ensure that the optimized wing is structurally sound}.

The optimization problem reads:
\begin{equation}
\label{eq:opt}
    \begin{split}
        \min_{\left\{\gamma;\,d\right\}\in\mathbb{R}}\quad& D_\mathit{cruise}(d)\\
        \mathrm{s.t.}\quad& g_1=\frac{P+W(\gamma,d)-L_\mathit{cruise}(d)}{P}\leq0\\
        & g_2=\frac{C_\mathit{cruise}(\gamma,d)-\overline{C}_\mathit{cruise}}{\overline{C}_\mathit{cruise}}\leq0\\
        & g_3=\frac{C_\mathit{takeoff}(\gamma,d)-\overline{C}_\mathit{takeoff}}{\overline{C}_\mathit{takeoff}}\leq0\\
        &0\leq\gamma_i\leq1\quad\forall i=1,2,\dots N_\mathit{struct}\\
        &0\leq d_i\leq1\quad\forall i=1,2,\dots N_\mathit{shape}
    \end{split}
\end{equation}
where box constraints are placed on both the structural variables, $\gamma$, and on the shape variables, $d$. For each airfoil section, $d$ contains one entry for each optimized shape parameter, which is retrieved using the mapping \eqref{eq:shapeinterp}.

Note that the compliance constraint, apart from its direct physical motivation, also is a way of enforcing small deformations of the wing, ensuring that the aerodynamic response computed on the undeformed wing is accurate. Compliance constraints are used instead of stress constraints, which would be more intuitive for aerospace applications. However, partly for simplicity and partly since the design domain is relatively open and smooth, significant design differences are not expected between a stress\ADD{-}\RM{ based} or compliance\ADD{-}based solution. An advantage of using compliance constraints is the reduced computational cost of the sensitivity analysis due to the system being self\ADD{-}adjoint (\cite{Bendsoe2003}). 

The Method of Moving Asymptotes (MMA) \cite{Svanberg1987} implemented in PETSc (\cite{Aage2013}) is used as optimization algorithm. Considering the different natures of the shape and topology optimization variables, move limits of 0.1 are applied to the structural design variables and 0.05 to the shape variables, respectively. Furthermore, conservative asymptote settings are used (increase 1.05 and decrease 0.65). Further, to ensure a well\ADD{-}conditioned problem for MMA, the drag is normalized by a tenth of its value at the first design iteration.

The $\beta_\mathit{skin}$ parameter used for the erosion-dilation process leading to the fixed wing skin thickness is constant throughout the optimization. The projection sharpness applied on the internal design variable $\gamma$, as described in \cref{sec:internal} is continued from an almost linear interpolation to an increasingly sharper Heaviside approximation. The optimization is initialized with $\beta=0.01$, leading to a quasi-linear interpolation. The optimization then runs until the design is close to feasibility i.e.\ADD{,} $\max(g_j)\leq0.05$, from where the continuation starts as seen in \cref{tab:cont}.
\begin{table*}[htb]
    \centering
    \caption{Continuation strategy applied on the Heaviside projection sharpness $\beta$ for the structural design.}
    \label{tab:cont}
    \begin{tabular}{l|rrrrrrrrrr}
        Iterations after initial feasibility & 0 & 5 & 45 & 85 & 115 & 145 & 175 & 205 & 235 & 265 \\
        $\beta$ & $0.01$ & $1$ & $2$ & $3$ & $4$ & $5$ & $6$ & $7$ & $8$ & $16$\\
        Payload $\mathrm{[kN]}$ & $9.52$ & $9.04$ & $8.60$ &  $8.17$ & $7.75$ & $7.37$ & $7.00$ & $6.65$ & $6.32$ & $6.00$
    \end{tabular}
\end{table*}

To accelerate the convergence towards a feasible design at each continuation step, the lift-weight-payload constraint, $g_1$, is overly constrained by starting the optimization process with an increased payload and subsequently relaxing it by $5\%$ at each continuation step shown in \cref{tab:cont}, such that the payload obtained after the last continuation step corresponds to the target.

\subsection{Sensitivity analysis}

The sensitivities, with respect to both shape- and topology design variables, of the objective- and three constraint functions seen in \eqref{eq:opt} are obtained using a coupled adjoint method. In the general case, the augmented Lagrangian function $\mathcal{F}$ of the function $f$ is considered. The residuals of the panel method and Finite Element Method from \eqref{eq:aerostate} and \eqref{eq:femRes}, respectively, can be added to a functional $f$, multiplied with their respective arbitrary Lagrangian multipliers, $\bm{\lambda}_a$ and $\bm{\lambda}_s$:
\begin{equation}
\label{eq:auglag}
    \mathcal{F} = f + \bm{\lambda}_a^\intercal\mathbf{r}_a+\bm{\lambda}_s^\intercal\mathbf{r}_s
\end{equation}
By differentiating the augmented Lagrangian function and reordering the terms, the one-way coupled adjoint equations can be set up, and solved sequentially:
\begin{equation}
\label{eq:adjeq}
\begin{split}
\left(\frac{\partial\mathbf{r}_s}{\partial\mathbf{u}}\right)^\intercal\bm{\lambda}_s&=\left(-\frac{\partial f}{\partial\mathbf{u}}\right)^\intercal\\
\left(\frac{\partial\mathbf{r}_a}{\partial\bm{\mu}}\right)^\intercal\bm{\lambda}_a&=-\left(\frac{\partial f}{\partial\bm{\mu}}\right)^\intercal-\left(\frac{\partial\mathbf{r}_s}{\partial\bm{\mu}}\right)^\intercal\bm{\lambda}_s
\end{split}    
\end{equation}
The sensitivities are then retrieved by:
\begin{equation}
\label{eq:sens}
\begin{split}
    \frac{d\mathcal{F}}{d\gamma} =& \frac{\partial f}{\partial \gamma}+\bm{\lambda}_s^\intercal\frac{\partial\mathbf{r}_s}{\partial\gamma}\\
    \frac{d\mathcal{F}}{dd^s_i} =& \frac{\partial f}{\partial d}+\bm{\lambda}_a^\intercal\frac{\partial\mathbf{r}_a}{\partial d^s_i}+\bm{\lambda}_s^\intercal\frac{\partial\mathbf{r}_s}{\partial d^s_i}
    \end{split}
\end{equation}

It is noted that when the objective functional is the compliance function, $f=C$, the structural adjoint is retrieved as the negative displacement field, $\bm{\lambda}_s=-\mathbf{u}$.

The chain rule term for the inside fraction of an element, with respect to panel coordinates (which subsequently are transformed into the shape design variables), is obtained using a central difference scheme. For each intersected element, the coordinates of the nodes connected to panels intersecting the element are perturbed in both directions and the sensitivity is computed based on the corresponding inside fractions, as can be done for XFEM methods (\cite{Sharma2017}).

Further details on the derivation of the sensitivities and chain rule terms are provided in \cref{app:sens}.

As the drag objective functional and the lift-weight constraint do not depend on the structural state, only the aerodynamic adjoint analysis \ADD{needs} to be carried out for these two functions. 

For the compliance constraints, the full one-way coupled adjoint system needs to be considered. The structural adjoint is first projected back to the panel basis using the load transfer scheme, which was used for the mapping of the pressures to the finite element mesh, as described in \cref{sec:loadtrans}. The structural term going into the shape sensitivity in \cref{eq:sens} can be decomposed into two terms;
\begin{equation}\bm{\lambda}_s^\intercal\frac{\partial \mathbf{r}_s}{\partial d^s_i}=\bm{\lambda}_s^\intercal\left(\frac{\partial\mathbf{K}}{\partial d^s_i}\mathbf{u}-\frac{\partial\mathbf{p}}{\partial d^s_i}\right)\end{equation}
where the pressure mapping derivative with respect to the wing shape is neglected - $\frac{\partial\mathbf{p}}{\partial d^s_i}\approx0$, as it is small but seems extremely sensitive to the current cut location. \ADD{These instabilities might be linked to the load transfer scheme only being piecewise continuous, with potential discontinuities appearing when the set of intersected elements is changed.} Including the contribution leads to jumps in compliance and convergence issues. A similar approach was taken by \cite{Picelli2017} for density\ADD{-}based fluid-structure interaction topology optimization. \ADD{A solution for alleviating these discontinuities could be to smooth the load transfer scheme by filtering the transferred loads, e.g. by using the PDE filter (\cite{Lazarov2011}). Filtering the loads would however result in less accurate modeling of the wing response. Furthermore, if the loads are filtered onto degrees of freedom belonging to a wider band of elements, a thicker wing skin would also be required to ensure that the loads are not projected onto degrees of freedom belonging only to void elements.}

\section{Results}
\label{sec:res}

\begin{figure}[htb]
    \centering
    \includegraphics[width=\linewidth]{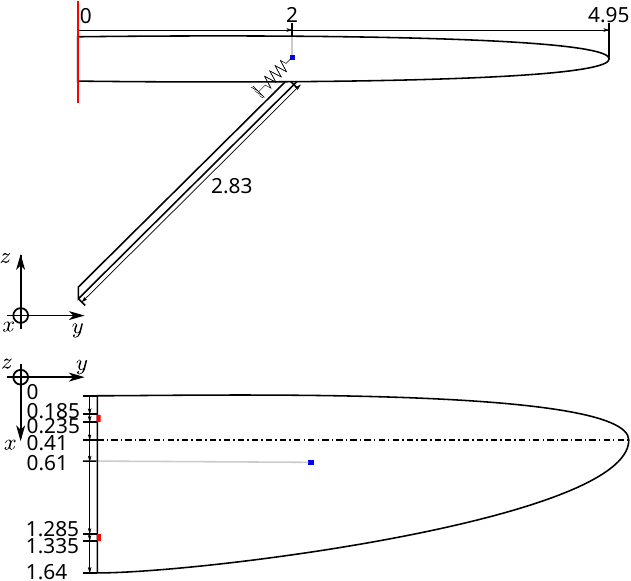}
    \caption{Sketch of the wing. The red bars in the planform view indicate the areas where the wing is clamped. The dashed line indicates the wing quarter-chord. Note that the strut is included for visualization purposes only. Instead, springs are added to the degrees of freedom in the blue box, as illustrated in the insert.}
    \label{fig:cessnaprob}
\end{figure}

The wing of a lightweight single-engine airplane inspired by a Cessna 172 is considered. The wing consists of a NACA 2412 airfoil with a  half span of $4.95\;\mathrm{m}$. Clamped boundary conditions are enforced at two locations at the root of the wing - one near the leading edge and one near the trailing edge - as marked in red on \cref{fig:cessnaprob}. As the wing at the root can be rotated by the optimization, the boundary conditions are enforced on the entire strips along the $z$ axis at the marked positions. 

A strut supporting the wing, as shown on the sketch in \cref{fig:cessnaprob}, is also considered in some cases. The strut is modeled by adding springs to the corresponding degrees of freedom in a volume, where the strut would be fixed to the wing. The fixation volume \ADD{is} a cube of side length $0.05\;\mathrm{m}$ centered behind the quarter chord position $y=2\;\mathrm{m}$. This volume is also depicted in \cref{fig:cessnaprob}.

The total stiffness of the springs representing the strut, $k$, is computed from the length of the strut, $l$, its cross sectional area, $A$ and the its Young's modulus, $E$:
\begin{equation}
k=\frac{AE}{l}    
\end{equation}

The wing is discretized using $40$ evenly spaced airfoil sections, which are each discretized into $100$ linear segments in the chordwise direction, where they follow a cosine distribution. The unfitted finite element mesh is a box of dimensions $1.68\times 4.96 \times 0.41\;\mathrm{m}$, discretized into $27\cdot 10^6$ hexahedral elements, with an edge length of $5\;\mathrm{mm}$.

For numerical reasons, all optimizations are run with normalized Young's modulus, i.e. $E=1\;\mathrm{Pa}$, and a Poisson's ratio of $\nu=0.3$. The contrast for the void material is set to $E_\mathit{min}/E_\mathit{max}=10^{-6}$. 

The optimization is initialized with a uniform wing with a twist of $3\degree$ and a $1.64\;\mathrm{m}$ chord length. The angle of attack may be varied from $-7\degree$ to $10\degree$ and the chord length from $0.1\;\mathrm{m}$ to $1.64\;\mathrm{m}$, in cases where it is allowed to vary. The camber is kept constant throughout the optimization. The shape variables are filtered with a radius of $\frac{1}{3}$ of the wing half-span, as suggested by \cite{Conlan-Smith2020}, to avoid the optimization exploiting numerical artifacts of the panel method. 

The structural design variable is initialized to $\gamma=0.5$ throughout the structural mesh, both inside and outside the wing. This means that elements that enter the wing will have intermediate values. The sensitivity of objective and constraints with respect to $\gamma$ outside the vicinity of the wing is zero, so the $\gamma=0.5$ is maintained until the element gets within a filter radius of the wing. The filter radius for the skin erosion-dilation process is set to $r_e=7.5\;\mathrm{mm}$ leading to a fixed wing-skin thickness of $10\;\mathrm{mm}$, corresponding to $2$ elements - thus ensuring that the aerodynamic forces are mapped to solid elements. Furthermore, the filter radius for the internal design of the wing is set to $r_s=12.5\;\mathrm{mm}$. The threshold value gap for the internal variables is set to $\Delta\eta=0.2$, leading to a minimum length scale of $11.25\;\mathrm{mm}$.
 
The relative wing skin stiffness is set to $\rho_\mathit{skin}=0.05$. This low stiffness for skin elements is chosen to reflect that the enforced wing skin thickness, is of $10\;\mathrm{mm}$ (which cannot be lower due to the discretization level), which is much higher, than what would normally be used on an airplane wing. \ADD{A high projection sharpness is chosen for the skin identification, $\beta_\mathit{skin}=64$, this ensures that the skin is sharply defined and of the chosen thickness.}
  
The angles of attack for the takeoff and cruise conditions are set to $8.5\degree$ and $4.5\degree$, respectively. Considering the \ADD{cross-section} area of the strut to be $A=5\cdot10^{-3}\;\mathrm{m^2}$ and its length $l=2.83\;\mathrm{m}$, \ADD{the strut stiffness is set to to $k=123\cdot 10^6\;\mathrm{N/ m}$, if considering aluminum as the strut material. As the computational Young's modulus is set to $\overline{E}=1\;\mathrm{Pa}$, the computational spring stiffness is set to $\overline{k}=k\overline{E}/E_\mathit{Alu}=1.76\cdot10^{-3}\;\mathrm{N/m}$.} The density of the material is set to $2.71\cdot 10^3\;\mathrm{kg/m^{3}}$ - corresponding to a density of $W_\mathit{structure}=26.59\cdot10^3\;\mathrm{N/m^{3}}$.

\ADD{The air density is $\rho_\mathit{air}=1.225\;\mathrm{kg/m^3}$ and the air speed is set to $V_\infty=70\;\mathrm{m/s}$ under both the cruise and takeoff conditions.}

The optimization is carried out on the DTU Sophia cluster (\cite{sophia}), using 10 nodes with 2 AMD EPYC 7351 16-core processors each, giving a total of 320 cores. One node is dedicated to the aerodynamic problem, while the remaining 9 are dedicated to the structural problem. Typically\ADD{,} the cost of a design iteration is between 350 and 650 seconds, averaging at 490 seconds. \ADD{Most} of the time is \ADD{spent} on the solution of the linear elasticity equations.

In total, four cases are considered. Two of the cases include the wing strut, as previously discussed and in the other two cases, the strut is omitted. For both cases with and without the strut, two optimizations are run, with different freedom in the shape part of the optimization; one case where only the twist is considered, and the other case with both twist and chord length.

\subsection{Optimization with strut}
The support of the strut is placed on the wing to optimize the wing of a small single-engine airplane, with struts under the wings. This strut provides some support away from the root of the wing, hence providing some resistance toward bending moments. It is expected that this strut plays a key role in the structural response and hence also in the obtained optimized structures, as well as in the optimized shapes and lift and drag distributions. However, for simplicity, the drag of the strut itself is not included in the optimization in this study. In reality\ADD{,} the strut would also interact with the wing aerodynamics. As the strut is relatively small, it is not deemed necessary to include it in the optimization.

The compliance constraints are set to $\overline{C}_\mathit{takeoff}=6\cdot 10^{11}\;\mathrm{J}$ and $\overline{C}_\mathit{cruise}=1.7\cdot10^{11}\;\mathrm{J}$, respectively. The compliance constraints can be converted to  $\overline{C}_\mathit{takeoff}=8.6\;\mathrm{J}$ and $\overline{C}_\mathit{cruise}=2.42\;\mathrm{J}$, respectively, when substituting the computational Young's modulus to the one of aluminum, $E=70\;\mathrm{GPa}$.  The compliance constraint values correspond to $\sim16\%$ of the compliance of the initial wing in the cruise conditions and $\sim30\%$ in the takeoff conditions.
\subsubsection{Optimization of twist and structure}
\begin{figure}[htb]
    \centering
    \includegraphics[width=\linewidth]{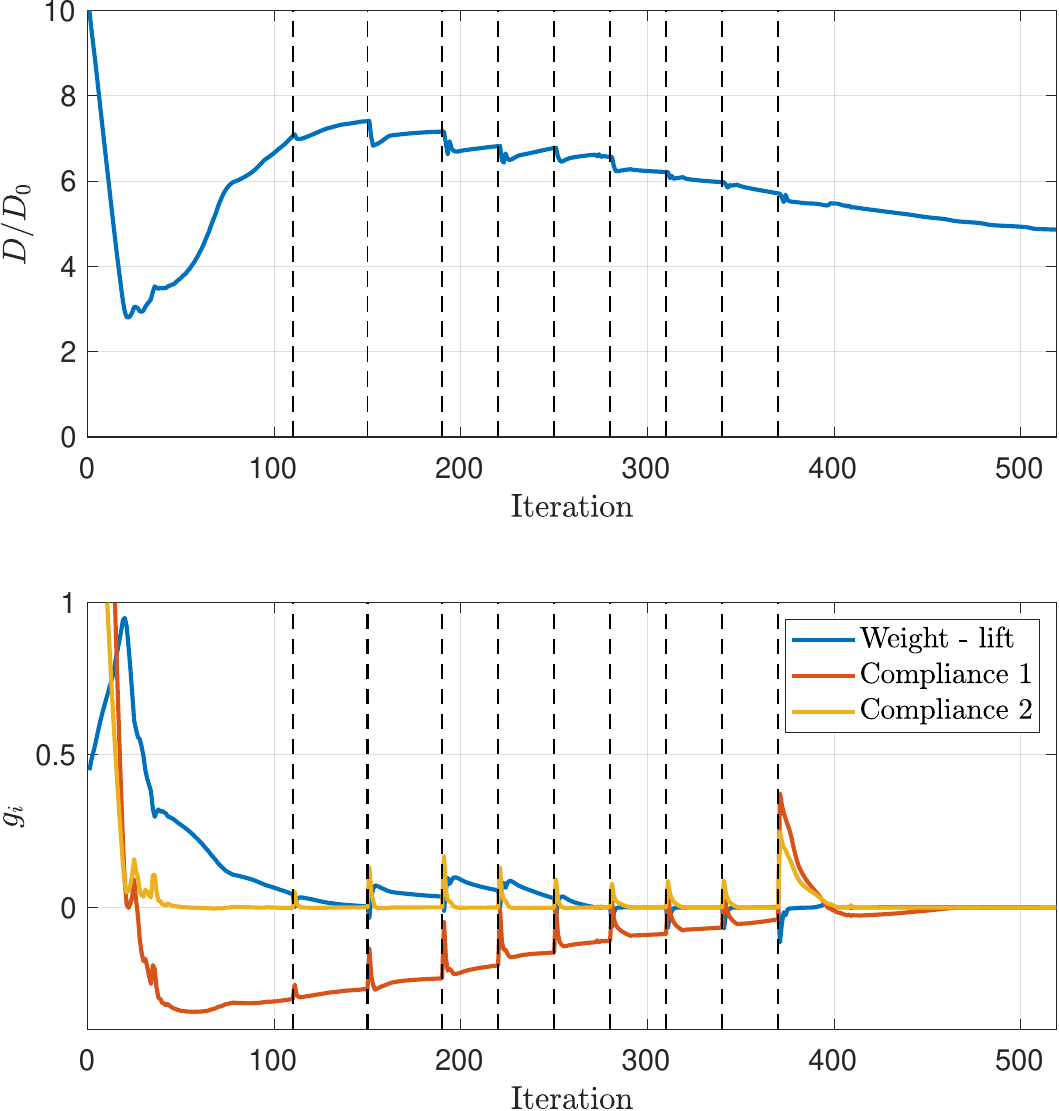}
    \caption{Optimization history of normalized objective and constraints for the wing optimized with the strut and only varying twist and internal structure. Dashed vertical lines indicate $\beta$ continuation steps.}
    \label{fig:strut_twist_hist}
\end{figure}

To showcase the presented methodology, the wing is first optimized by only varying the wing shape through the local twist. The optimization is run for a total of 450 iterations. The  feasibility criteria for the start of the optimization, i.e., $g_i\leq0.05$ is achieved after 107 iterations. The optimization history is seen in \cref{fig:strut_twist_hist}.

The initial part of the optimization process mostly focuses on achieving feasibility on the design. Here it is especially noted that the objective and constraint functions are conflicting: For the compliance constraints extensive internal structure and low aerodynamic forces are desired. For the weight-lift constraint, a low mass of internal structure and high lift (i.e. high aerodynamic forces) are desired. Finally, the objective deals with minimizing the drag - which also means reducing the lift.

The bumps in objective and constraint functions in the optimization history coincide with the design iterations where the $\beta$ parameter for the internal design is \ADD{continued} using the scheme shown in \cref{tab:cont}.

\begin{figure}[htb]
    \centering
    \includegraphics[width=\linewidth]{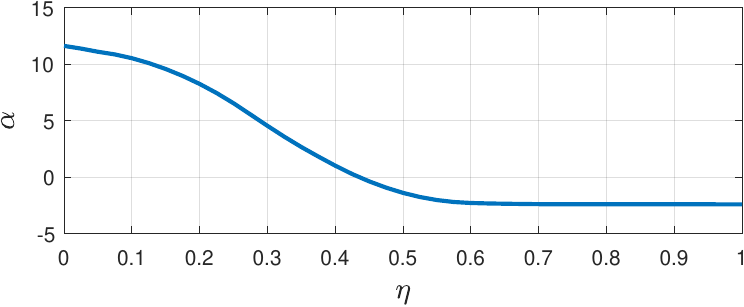}
    \caption{Cruise angle of attack, $\alpha$ distribution as a function of the relative spanwise position $\eta$, of wing optimized with the strut and only optimizing the twist and internal structure.}
    \label{fig:strut_twist_distrib}
\end{figure}

The local angle of attack distribution at cruise conditions is shown in \cref{fig:strut_twist_distrib}. Here it is observed that the highest twist is present at the root of the wing and decreases to the minimum around the half span. Keeping in mind that the strut spring locations are placed at a spanwise position of $\eta=0.4$, it is noted that the twist is reduced to the minimum beyond that position. As the airfoil in the wing is non-symmetric, the negative angle of attack does not necessarily contribute negatively to the lift, as shown later. It is noted that the Reynolds number at the investigated flow regime is approximately $\mathit{Re}\approx8\cdot10^6$. For such Reynolds numbers ($\mathit{Re}=9\cdot10^6$), \cite{Abbott1945} observed that flow separations on NACA 2412 airfoil sections occur at angles of attack $\alpha\geq16\degree$. It is hence seen that flow separation should not be an issue for the angles of attack shown in \cref{fig:strut_twist_distrib}, the panel method should therefore be able to provide accurate solutions in the investigated configuration.

\begin{figure}[htb]
    \centering
    \includegraphics[width=\linewidth]{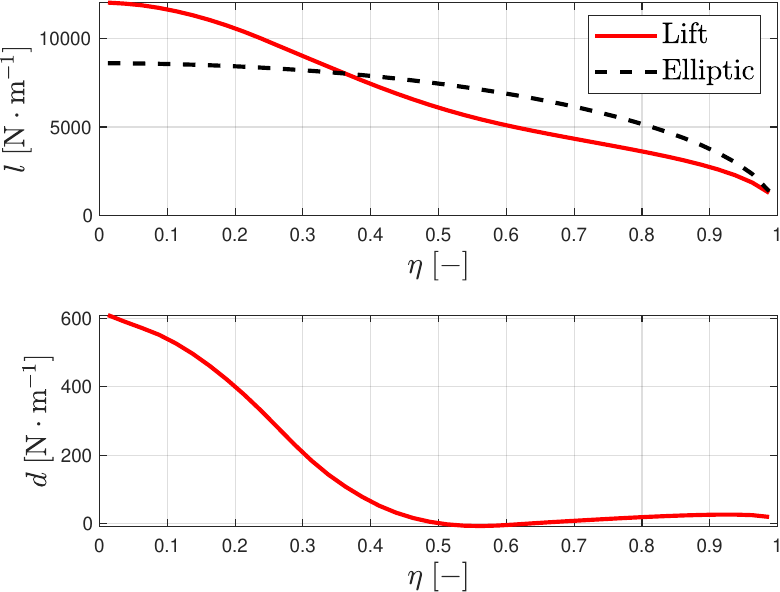}
    \caption{Lift, $l$, and drag, $d$, distributions along the spanwise position $\eta$ evaluated on the Trefftz plane of the wing optimized with the strut and only varying the twist and internal structure.}
    \label{fig:strut_twist_lift}
\end{figure}

The lift and drag distributions along the spanwise position are shown in \cref{fig:strut_twist_lift}. Here it is seen that the lift distribution is close to the one observed by \cite{Drela2014} for wings with a large span when a constraint is placed on the root bending moment of the wing. A difference is that the distribution obtained here has a kink around the normalized span position $\eta=0.45$, which is close to the strut. Similarly, it is observed that \ADD{most} of the drag is produced between the root of the wing and the strut mount. This is likely linked to the increased load-bearing capacity between the root of the wing and the strut. Hence, the main part of the load \ADD{is} placed on the wing between the root and the strut, as this area is subjected to some double-supported conditions: the two clamped conditions at the root and the strut.

\begin{figure*}[htb]
    \centering
   \begin{subfigure}{0.49\linewidth}
   \includegraphics[width=\linewidth]{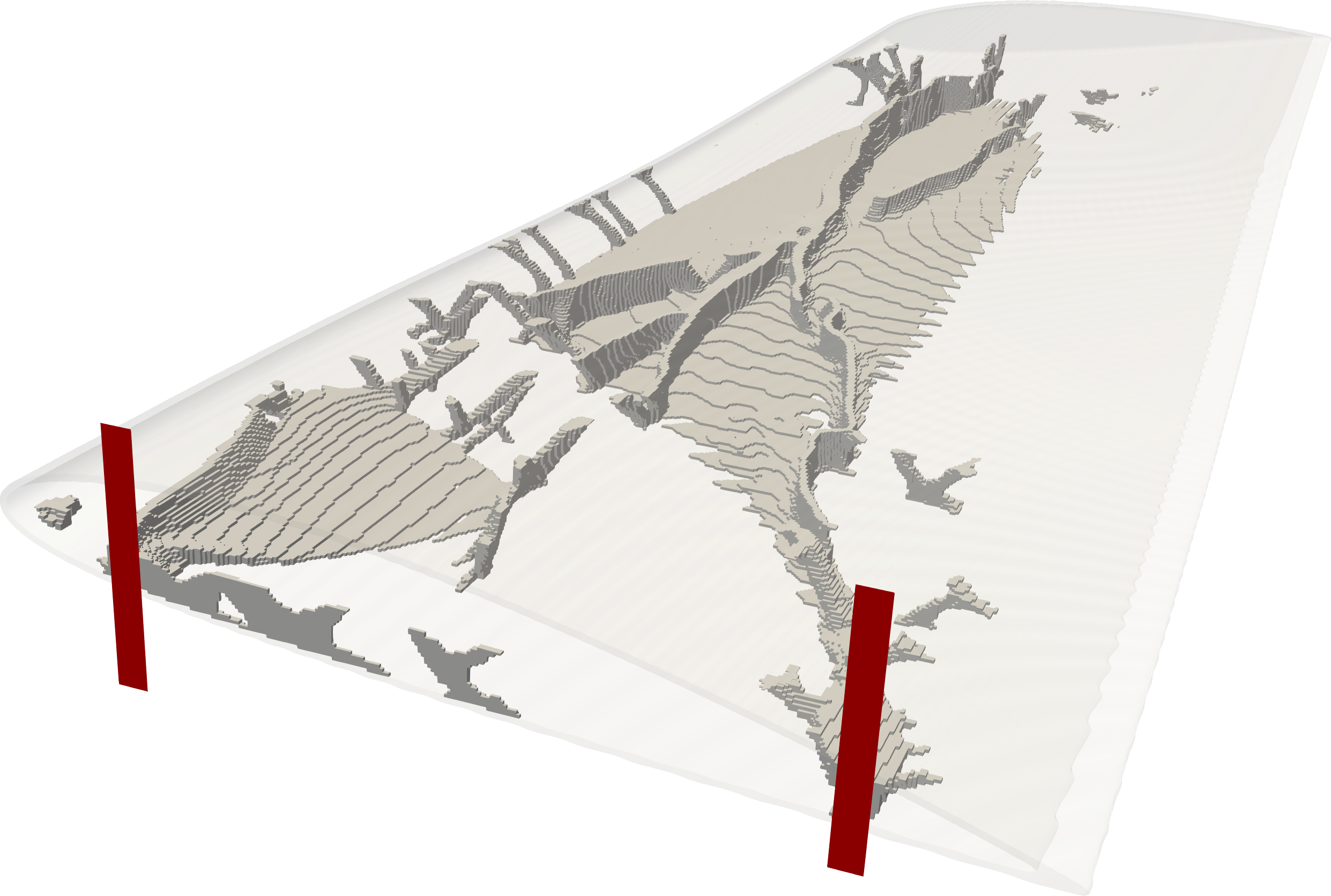}
   \caption{ }
   \end{subfigure}~
    \begin{subfigure}{0.49\linewidth}
    \includegraphics[width=\linewidth]{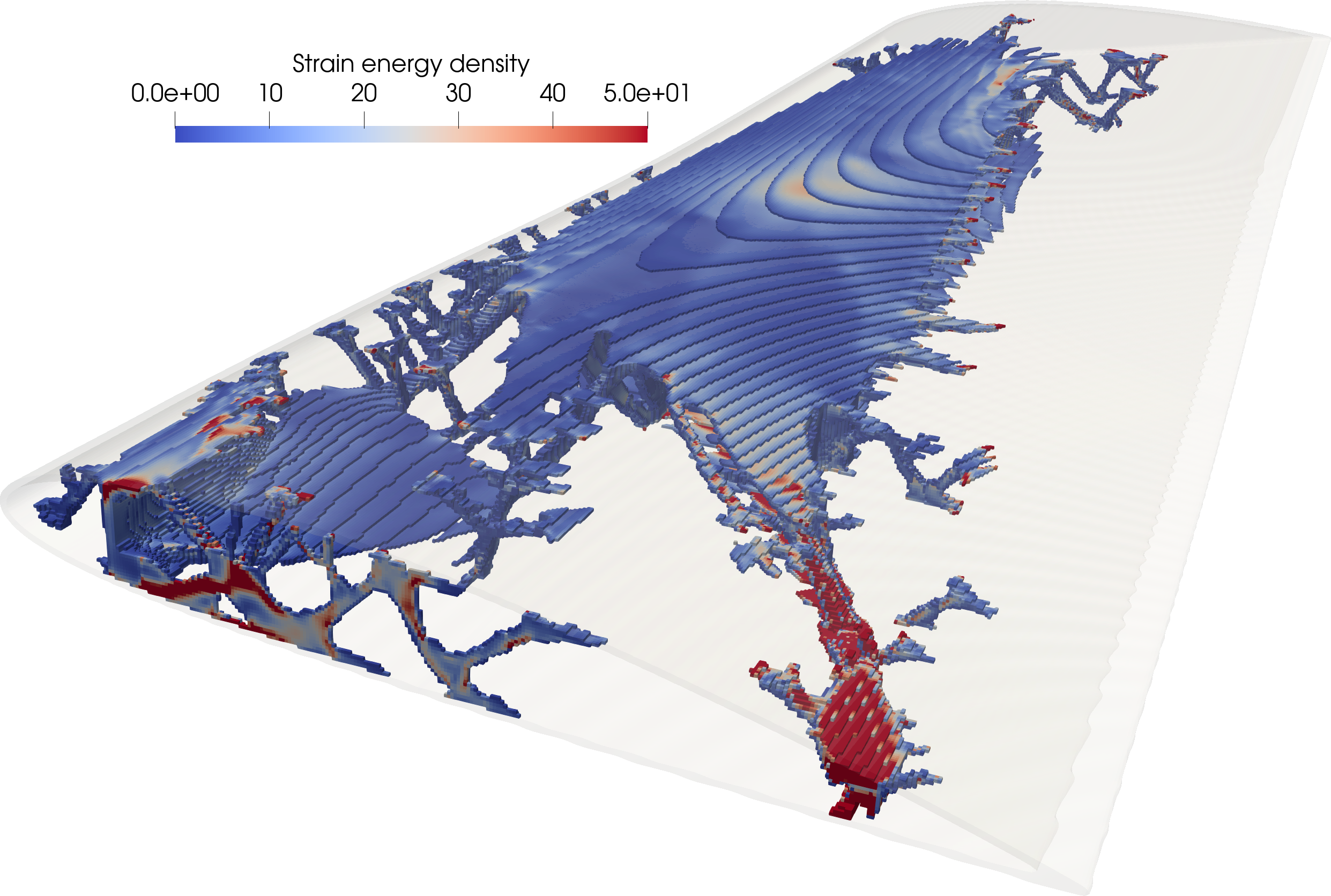}
    \caption{ }
    \end{subfigure}
    \caption{Structure of the wing optimized with the strut and only varying twist and internal structure. The figures show the fixed thickness skin in a transparent representation and (a) the sliced internal structure (i.e. only the lower part) and (b) the internal structure colored by strain energy density (adjusted for aluminum material properties) in $\mathrm{J\cdot m^{-3}}$. The regions with the clamped boundary conditions are depicted in red in (a).}
    \label{fig:strut_twist_struct}
\end{figure*}

The obtained internal structure of the optimized wing is shown in \cref{fig:strut_twist_struct}. Qualitatively, the design can be divided into two sub-areas: the locations between the root and the strut connection and the one between the strut connection and the tip of the wing. In both areas, the fixed thickness wing skin is reinforced with additional material around the quarter-chord location, this is most likely due to these locations being the furthest locations from the bending axis - hence providing high bending resistance. At locations between the root and the strut connection, multiple vertical stiffeners are placed and connect the skin to the internal structure. At locations further towards the tip, the joined spar branches out towards the tip again. A disconnection in the structure between the strut connection and supports is also observed. It should be noted that there is a line between the strut and the root of the wing, where there is no bending moment. This explains the lack of material inside the wing at the locations, which are in the vicinity of this line, as the skin provides enough stiffness in these locations. A full connection could be achieved by adding more \ADD{load cases}, as the point with no bending moment would be shifted across \ADD{load cases}, hence requiring an even distribution of material.

The rib, which can be seen at the root of the wing, stiffens the skin between the two fixtures at the root, which also is the location where the forces acting on the wing are the greatest. The rib could also have the role of transferring the forces in the skin to the two supports.

The strain energy density distribution in the internal structure of the wing, which is shown in \cref{fig:strut_twist_struct}, is evenly distributed throughout the structure, \ADD{except for} the region near the supports, where the energy density is higher.

\ADD{The final displacements of the wing are computed, to ensure that they are indeed small in the final optimized wing. When corrected for the aluminum Young's modulus, the maximum displacement magnitudes in the skin are found as $\max\left(\left|u_\mathit{cruise}\right|\right)=1.51\;\mathrm{mm}$ and $\max\left(\left|u_\mathit{takeoff}\right|\right)=3.35\;\mathrm{mm}$, under cruise and takeoff conditions respectively, which is below $0.1\%$ of the wing span and thus satisfying the assumption of small displacements.}

\subsubsection{Optimization of chord length, twist, and structure}
A second optimization case is run, where both the local twist and the chord length of the wing shape can be varied. The optimization is started with the same initial design as previously, and the continuation strategy is activated once $g_i\leq0.05$. 

\begin{figure}[htb]
    \centering
    \includegraphics[width=\linewidth]{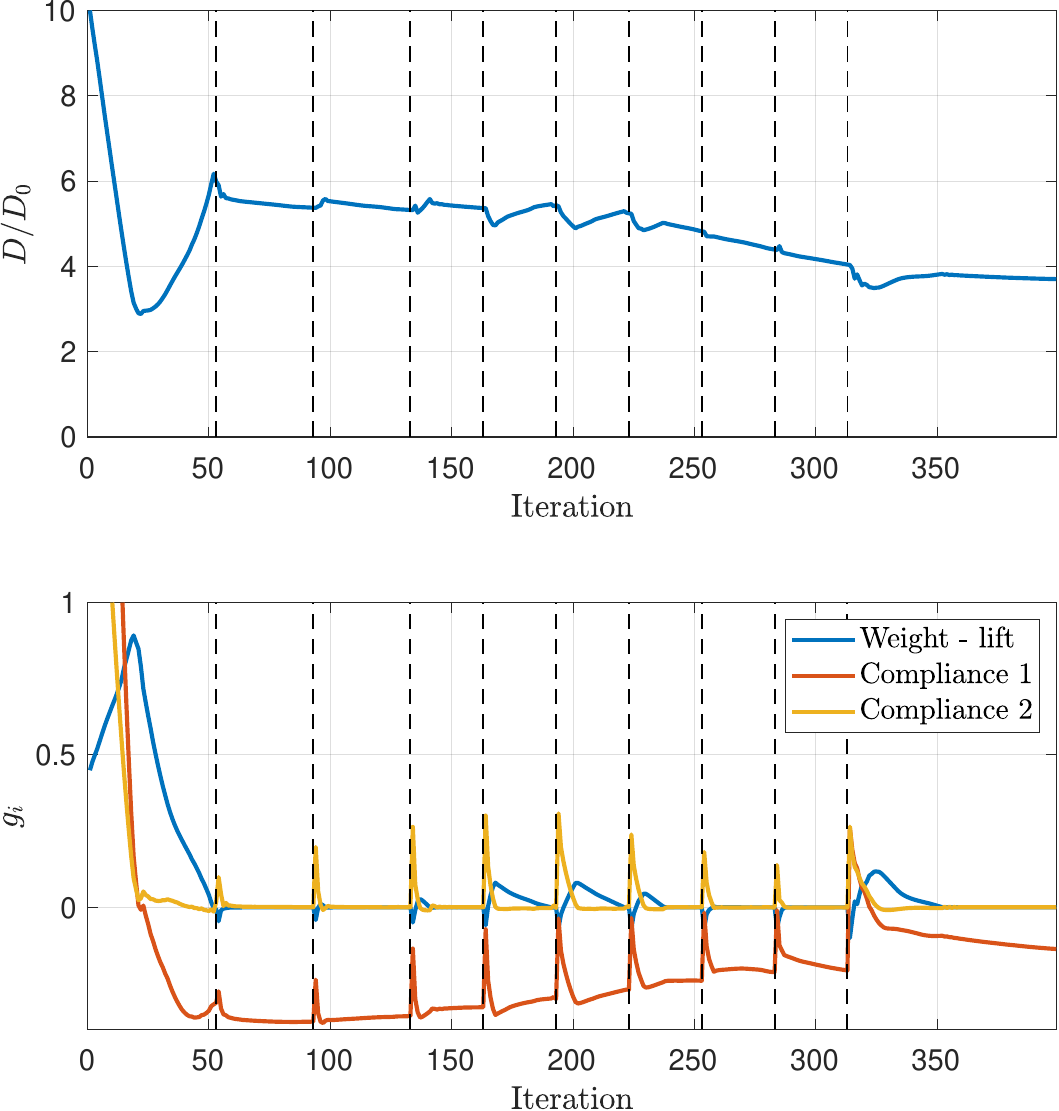}
    \caption{Optimization history of normalized objective and constraints for wing optimized with the strut and varying both local twist and chord length, as well as the internal structure. Dashed vertical lines indicate $\beta$ continuation steps.}
    \label{fig:strut_twist_chord_hist}
\end{figure}

From the optimization history, seen in \cref{fig:strut_twist_chord_hist}, it is seen that a feasible design is reached much quicker than previously. This is attributed to the increased design freedom. Furthermore\ADD{,} it is noted that the final cruise drag (objective function) is $31\%$ lower than for the case where the wing shape is optimized using only the twist. This is both due to the higher freedom in the aerodynamic optimization, and due to the increased ability to reduce the weight of the fixed wing skin by reducing the chord length - hence necessitating less lift, which allows for wings, which induce less drag.

\begin{figure}[htb]
    \centering
    \includegraphics[width=\linewidth]{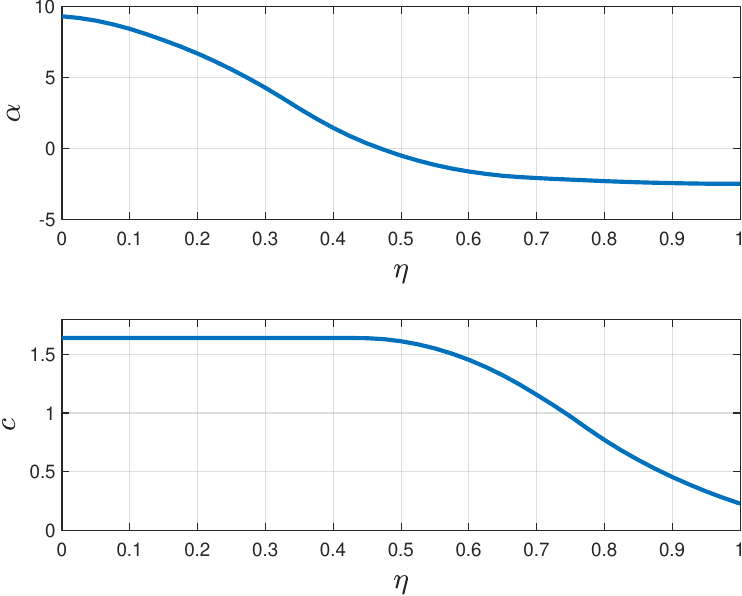}
    \caption{Local cruise angle of attack, $\alpha$, and chord length, $c$, distributions as a function of the relative spanwise position $\eta$, of wing optimized considering the strut and optimizing the local twist, chord length, and the internal structure.}
    \label{fig:strut_twist_chord_distrib}
\end{figure}

The cruise angle of attack and chord length distributions seen in \cref{fig:strut_twist_chord_distrib} confirm the previous results, when only optimizing the shape by means of the local twist. The amplitude of the variations is less than when only optimizing the twist. Furthermore, a connection between the twist and the chord length seems to exist; first, the twist is decreased to an almost constant low value outwards of the relative span position $\eta=0.6$. As the slope of the decrease in local twist decreases, the local chord length begins to drop from the maximum value near $\eta=0.5$ and reaches its minimum value at the tip of the wing. A higher chord length and lower twist might have been used closer toward the root if the bounds for the chord length were different. 

\begin{figure}[htb]
    \centering
    \includegraphics[width=\linewidth]{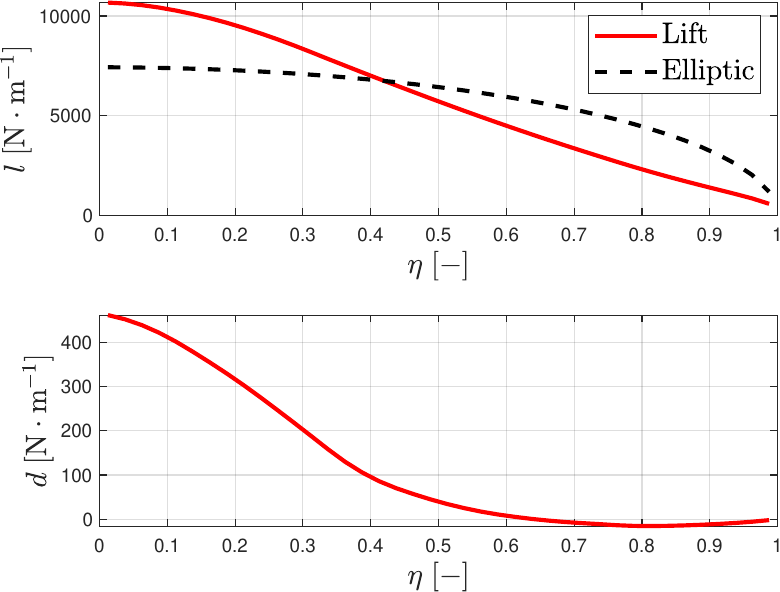}
    \caption{Lift and drag distributions of the wing optimized with the strut considering the local twist and chord, as well as the internal structure.}
    \label{fig:strut_twist_chord_lift}
\end{figure}

The lift and drag distributions along the span of the wing are shown in \cref{fig:strut_twist_chord_lift}. The distributions seem similar compared to the ones from the wing optimized with only the local twist and structure. However\ADD{,} the amplitude of the kink in the lift near the half-span location is decreased, which could indicate that the kink was present mainly for structural reasons. Similarities to the optimization with root bending moment constraint from \cite{Drela2014} are greater than when only optimizing the local twist, due to higher freedom in the aerodynamic part of optimization. The increased design freedom also leads to a better performance of the optimized design, which is also expected.

\begin{figure*}[htb]
    \centering
    \begin{subfigure}{0.45\linewidth}
    \includegraphics[width=\linewidth]{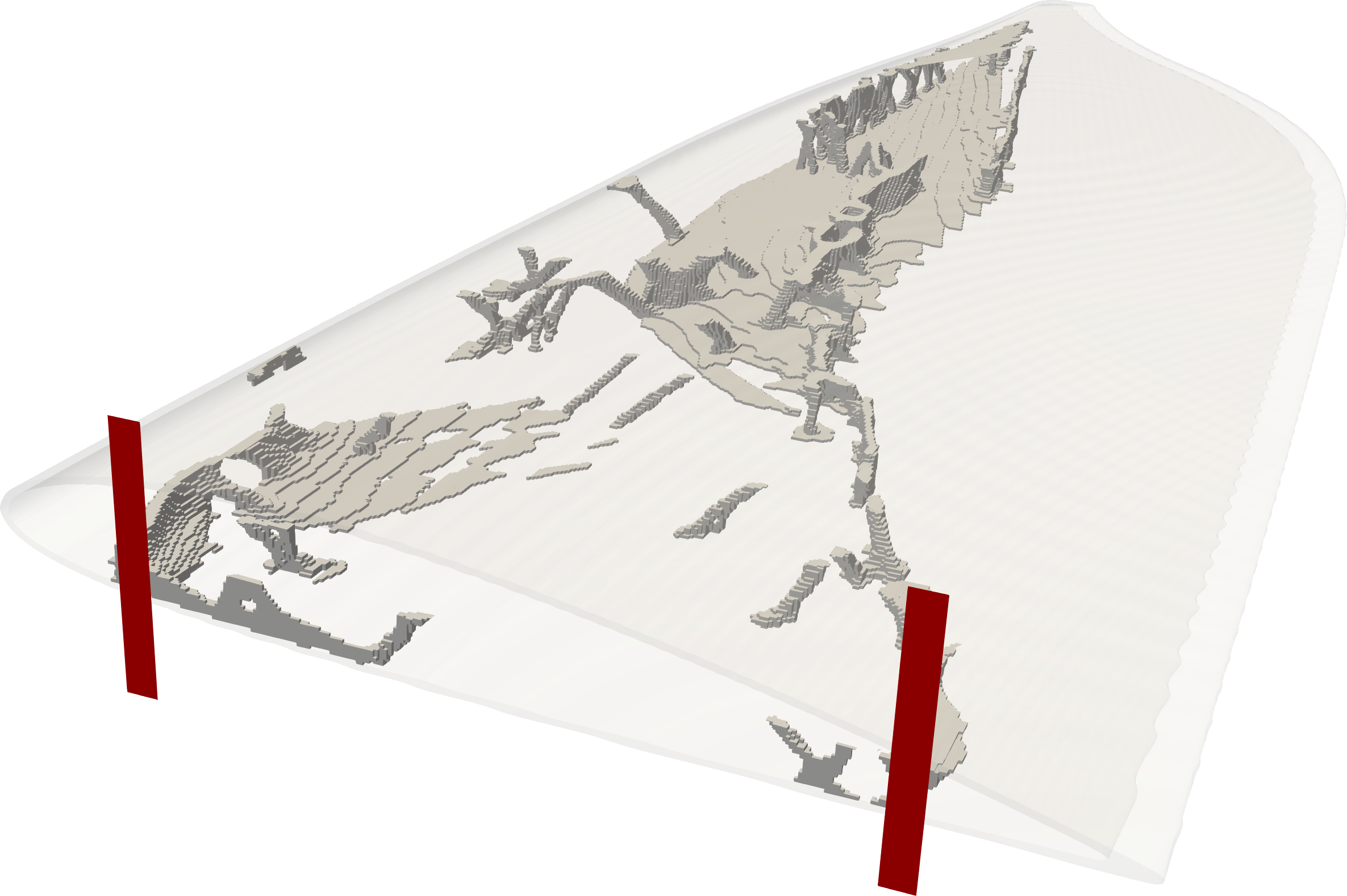}
    \caption{ }
    \end{subfigure}~
    \begin{subfigure}{0.45\linewidth}
    \includegraphics[width=\linewidth]{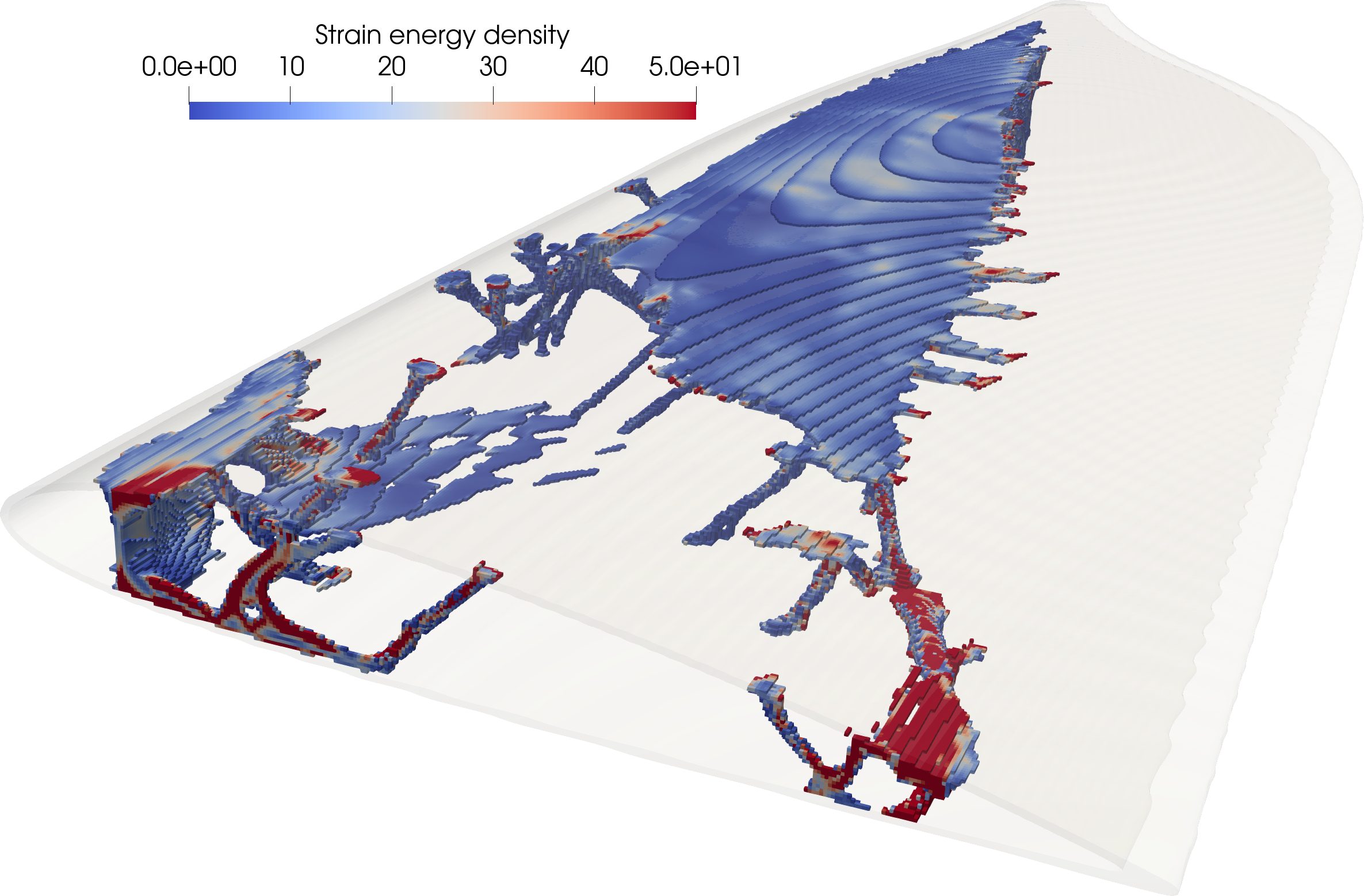}
    \caption{ }
    \end{subfigure}
     \caption{Structure of the wing optimized with the strut and varying twist, chord length, and internal structure. The figures show the fixed thickness skin in a transparent representation and (a) the sliced internal structure (i.e. only the lower part) and (b) the internal structure colored by strain energy density (adjusted for aluminum material properties) in $\mathrm{J\cdot m^{-3}}$. The regions with the clamped boundary conditions are depicted in red in (a).}
    \label{fig:strut_twist_chord_struct}
\end{figure*}

The obtained internal structure is shown in \cref{fig:strut_twist_chord_struct}. \ADD{Less} material is used in this case, \ADD{than} when a fixed chord length is used, which is due to the higher design freedom of the outer shape on the wing. As in the previous result, the structure is connected to the strut, indicating that a large portion of the internal forces is transferred to that part of the wing. However, less material seems to be connecting the strut-supported area to the root of the wing, which indicates that the skin is enough to carry loads of the low bending moments between the root and the strut. This could be due to the forces being transferred through the wing skin and the strut taking a higher part of the load. Structures looking like ribs are also observed on the internal structure adjacent to the top part of the wing skin. These ribs contribute to alleviating the pressure loads on the skin of the wing.

\ADD{As in the previous example, the displacements of the wing are monitored in the final design. When corrected for the aluminum Young's modulus, the maximum displacement magnitudes in the skin are found as $\max\left(\left|u_\mathit{cruise}\right|\right)=1.19\;\mathrm{mm}$ and $\max\left(\left|u_\mathit{takeoff}\right|\right)=5.26\;\mathrm{mm}$, under cruise and takeoff conditions respectively, which is below $0.11\%$ of the wing span.}

\subsection{Optimization without strut}
Two wings, with different amounts of design freedom for the external shape, are also optimized without considering the strut. Hence, the only considered structural supports are the two clamped conditions at the root of the wing, shown in \cref{fig:cessnaprob}. To ensure that a feasible point is reachable, the maximum compliances are increased to $\overline{C}_\mathit{takeoff}=120\cdot10^{11}\;\mathrm{J}$  and $\overline{C}_\mathit{cruise}=34\cdot10^{11}\;\mathrm{J}$, which can be converted to $\overline{C}_\mathit{takeoff}=172\;\mathrm{J}$ and $\overline{C}_\mathit{cruise}=48.4\;\mathrm{J}$, if the considered wing material is aluminum. These compliance constraints correspond to $\sim 20\%$ of the initial wing in cruise conditions and $\sim 36\%$ for takeoff conditions. When setting more restrictive upper bounds on the compliances, the optimization fails to reach a feasible point and gets stuck in a situation, where the two compliance constraints are feasible, but the lift is too low to overcome the weight-lift constraint and increasing it would lead to violations of the compliance constraints. 

Objective values for the two upcoming optimizations are not directly comparable to the optimizations carried out considering the strut, as the maximum compliances are different.
\subsubsection{Optimization of twist and structure}

\begin{figure}[htb]
    \centering
    \includegraphics[width=\linewidth]{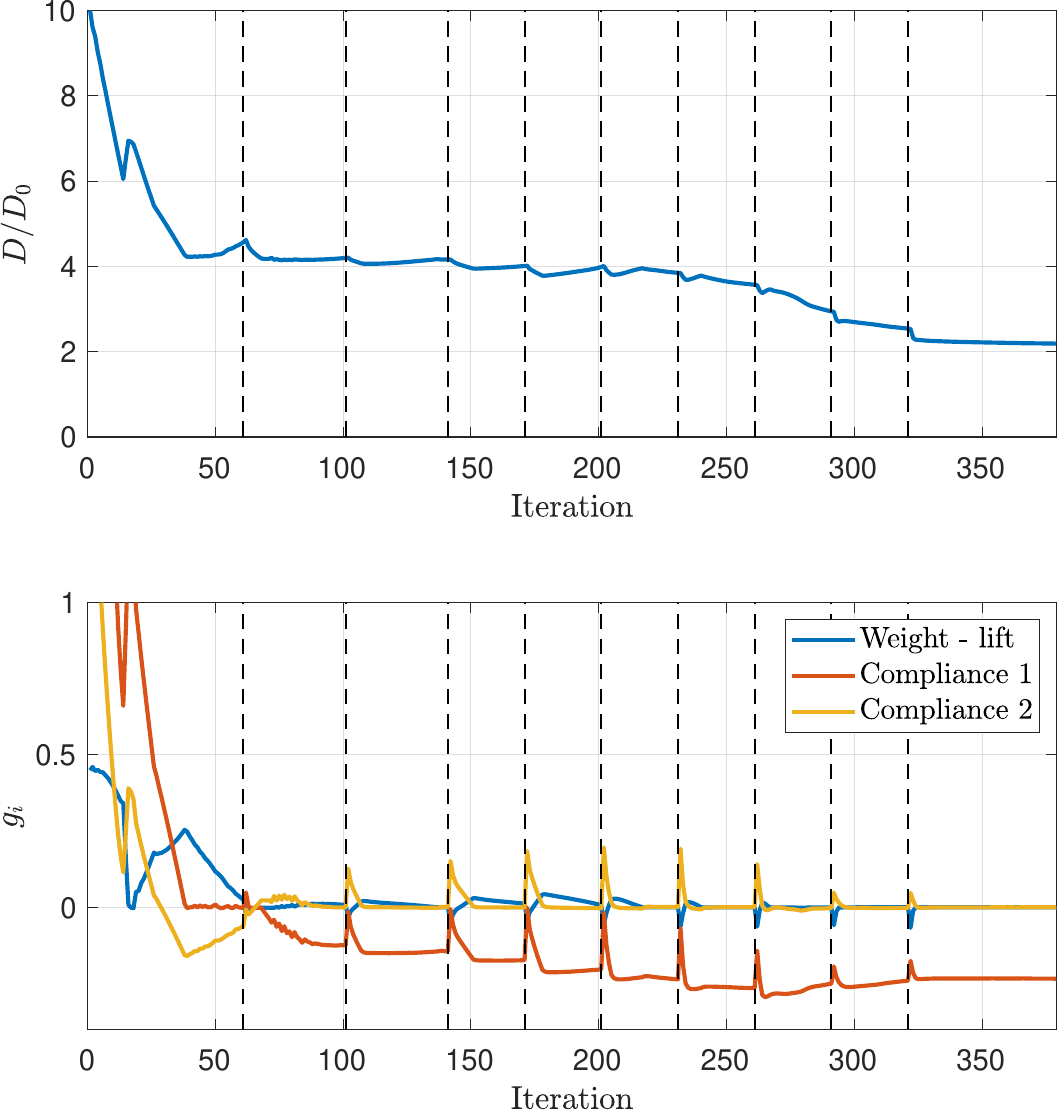}
    \caption{Optimization history of normalized objective and constraints for wing optimized without the strut and only varying local twist, as well as the internal structure. Dashed vertical lines indicate $\beta$ continuation steps.}
    \label{fig:nostrut_twist_hist}
\end{figure}

The optimization history of the wing optimized without the strut, considering only the local twist and internal structure is shown in \cref{fig:nostrut_twist_hist}. As in the designs optimized with the strut, the first part of the optimization is spent retrieving a feasible design from the initial one. A rise in $g_1$ simultaneous to decreasing values of the compliance constraints, $g_2$ and $g_3$, supports the hypothesis that decreasing the compliance constraints is partly achieved by reducing the aerodynamic forces on the wing - which results in decreased lift and drag.  

\begin{figure}[htb]
    \centering
    \includegraphics[width=\linewidth]{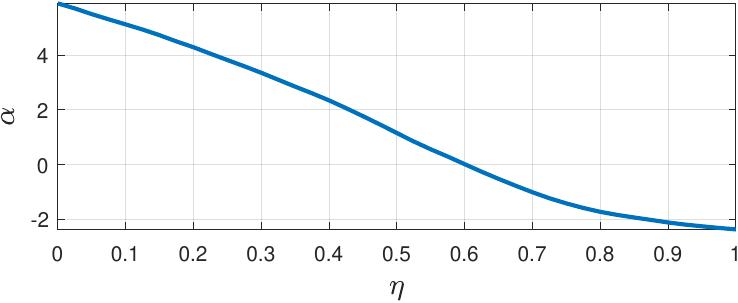}
    \caption{Cruise angle of attack, $\alpha$ distribution as a function of the relative spanwise position $\eta$, of wing optimized without the strut and only optimizing the twist and internal structure.}
    \label{fig:nostrut_twist_distrib}
\end{figure}

The local twist distribution of the optimized wing is shown in \cref{fig:nostrut_twist_distrib}, where it is noted that the local twist takes much more intermediate values than was seen for the wings optimized using the strut. This may be due to the compliance constraints being different (hence requiring less material leading to less lift required), and the lack of motivation to place the main part of the load between the wing root and the strut. The local twist flattens out toward the tip of the wing, which indicates that the load should be placed closer toward the root of the wing, which results in lower root bending moments.

\begin{figure}[htb]
    \centering
    \includegraphics[width=\linewidth]{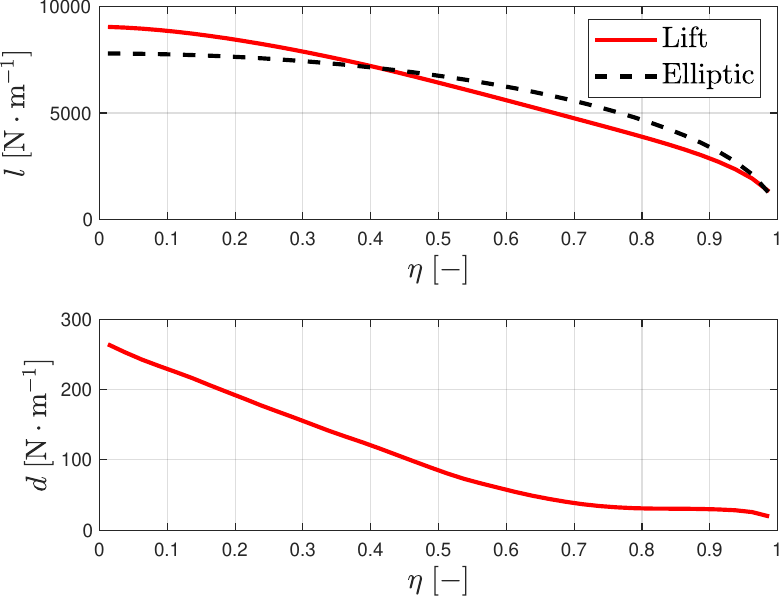}
    \caption{Lift, $l$, and drag, $d$, distributions of the wing optimized without the strut, considering the local twist and internal structure.}
    \label{fig:nostrut_twist_lift}
\end{figure}

The lift and drag distributions along the span \ADD{of the wing} are seen in \cref{fig:nostrut_twist_lift}. Here it is seen that the lift distribution is much closer to an elliptic profile than in the examples with the strut. However, the distribution has a higher part of the total load placed towards the root than the elliptic profile. This is also observed when enforcing a bending moment constraint at the root of the wing (\cite{Conlan-Smith2020}). It is amongst others observed that the drag seems to be at relatively high levels in the region near the tip of the wing than when considering the strut. This might be due to a larger portion of the lift being generated further away from the root. This is in turn caused by the absence of the strut and hence not being advantageous to produce large parts of the lift between the root and the strut location anymore.

\begin{figure*}[htb]
    \centering
    \begin{subfigure}{0.45\linewidth}
    \includegraphics[width=\linewidth]{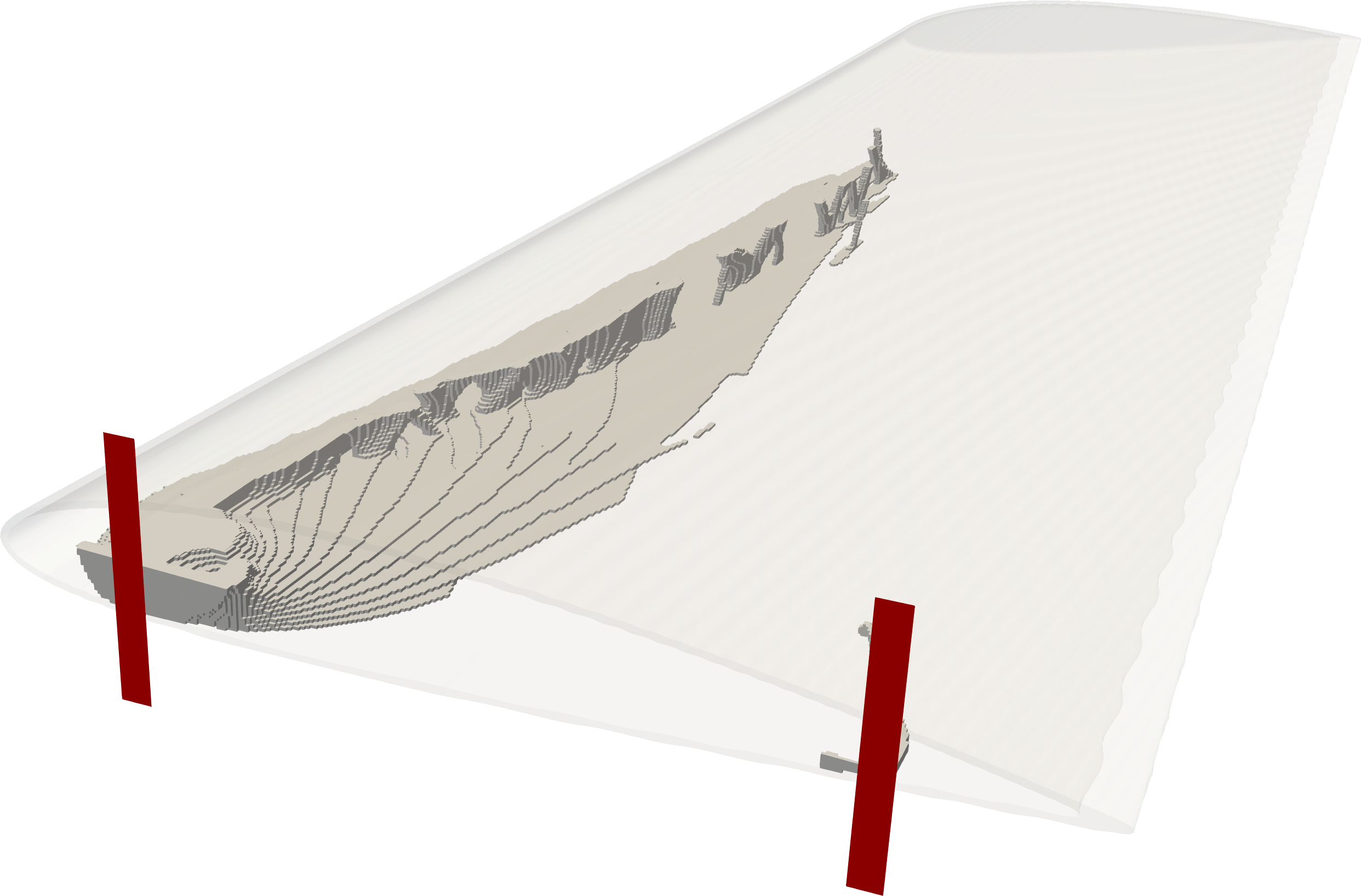}
    \caption{ }
    \end{subfigure}~
    \begin{subfigure}{0.45\linewidth}
    \includegraphics[width=\linewidth]{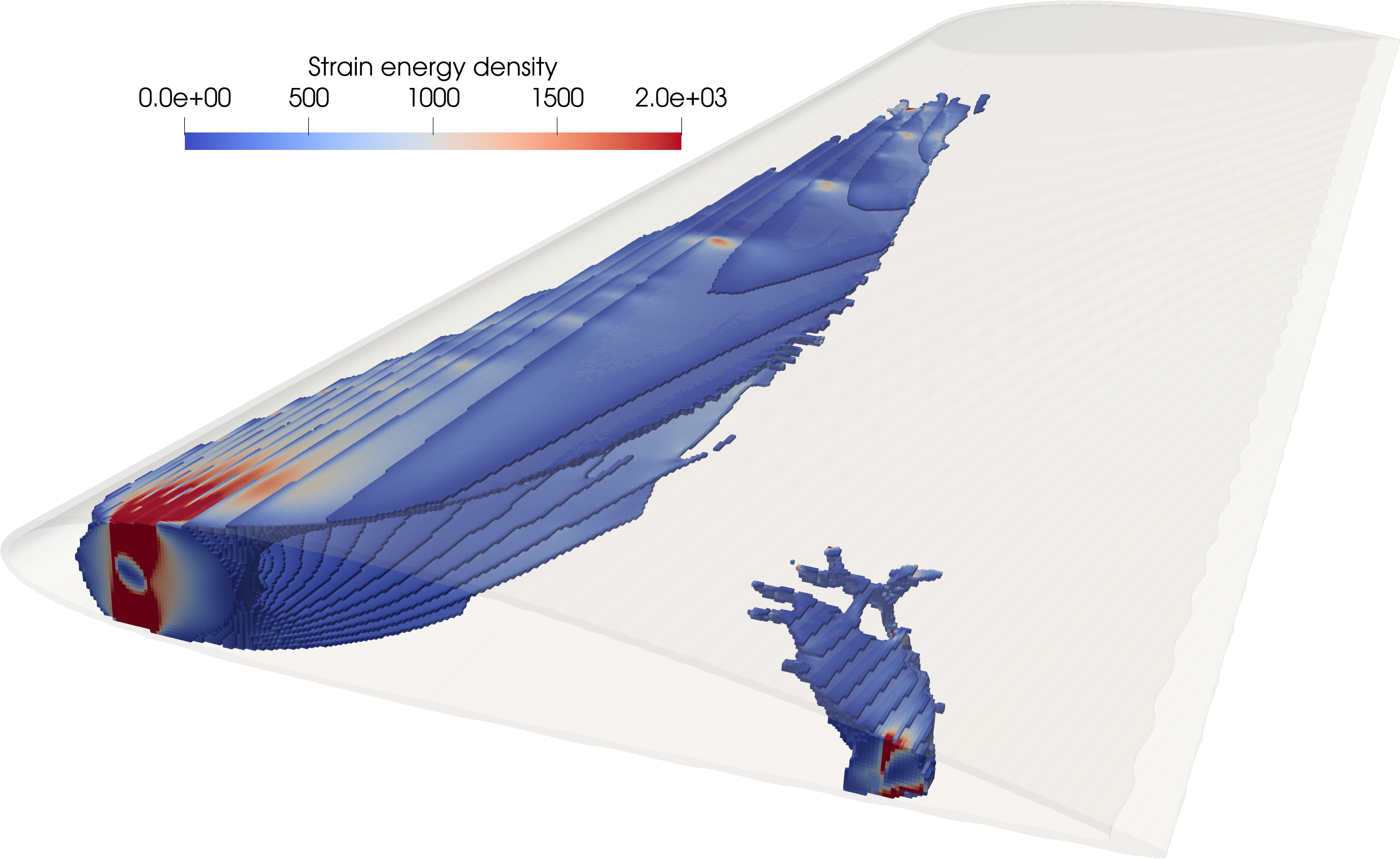}
    \caption{ }
    \end{subfigure}
     \caption{Structure of the wing optimized without the strut and only varying twist and internal structure. The figures show the fixed thickness skin in a transparent representation and (a) the sliced internal structure (i.e. only the lower part) and (b) the internal structure colored by strain energy density (adjusted for aluminum material properties) in $\mathrm{J\cdot m^{-3}}$. The regions with the clamped boundary conditions are depicted in red in (a).}
    \label{fig:nostrut_twist_struct}
\end{figure*}

The structure of the wing optimized without the strut, considering the local twist and internal structure, is seen in \cref{fig:nostrut_twist_struct}. The obtained structure is much different from the wing optimized with the strut from \cref{fig:strut_twist_struct}. The internal structure of the wing consists of an I-beam-like structure near the quarter chord line. The quarter chord line is also approximately the aerodynamic center of the wing, hence the derivative of the moment coefficient along that line is independent of the angle of attack, $\frac{dc_m}{d\alpha}=0$. The single strut is hence placed along the line where there are only low torsional contributions to the compliance and the primary load results from bending moments. The main structure is connected to both the upper and lower skin and the clamped support area near the leading edge. A much smaller secondary structure connects the upper and lower skin in a small area near the clamped support at the trailing edge of the wing. This secondary structure has the function of transferring torsional moments from the wing skin to the clamped support in the part of the wing, where the main beam is offset from the quarter chord, as it needs to connect to the clamped surface. The lack of other structure is due to the placement of the beam along the aerodynamic center, and there being no need to transfer forces from the positions close to the edges towards the aerodynamic center, as the skin seems to take that role in vast parts of the wing.

\ADD{The displacements of the wing are corrected for the aluminum Young's modulus and again reported in the final design. The maximum displacement magnitudes in the skin are found as $\max\left(\left|u_\mathit{cruise}\right|\right)=18.9\;\mathrm{mm}$ and $\max\left(\left|u_\mathit{takeoff}\right|\right)=41.3\;\mathrm{mm}$, under cruise and takeoff conditions respectively, which is below $0.9\%$ of the wing span.}

\subsubsection{Optimization of chord length, twist, and structure}

\begin{figure}[htb]
    \centering
    \includegraphics[width=\linewidth]{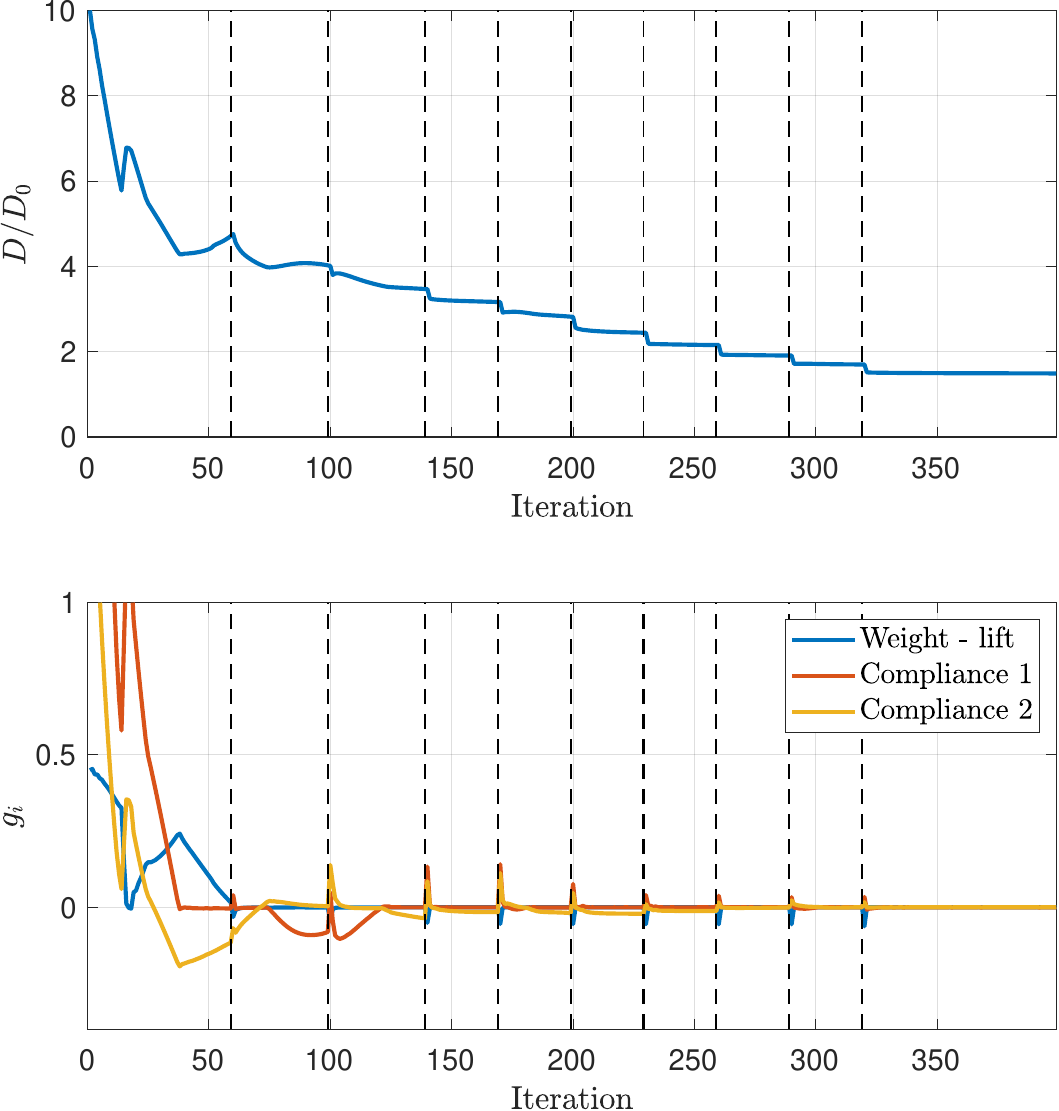}
    \caption{Optimization history of normalized objective and constraints for wing optimized without the strut and varying both local twist and chord length, as well as the internal structure. Dashed vertical lines indicate $\beta$ continuation steps.}
    \label{fig:nostrut_twist_chord_hist}
\end{figure}

The optimization history for the wing optimized without the strut, considering the local twist and chord length as well as the internal structure is shown in \cref{fig:nostrut_twist_chord_hist}. The optimization history shows that a feasible design is obtained much faster than in the previous case, which can be attributed to the increased design freedom. The required trade-offs between the structure and the aerodynamics to have all constraints feasible are again highlighted by the initial part of the optimization history as the lift-weight constraint, $g_1$, seems to counteract the compliance constraints, $g_2$ and $g_3$. The cruise drag of the final optimized design is $32\%$ lower than the one of the wing without the strut optimizing the twist and the structure. Contrary to what was observed in \cref{fig:nostrut_twist_hist}, with the history of the case with no strut and only optimization of the shape through the twist, both compliance constraints end up being active for most of the optimization, as well as in the final design. This can be attributed to the higher design freedom on the shape, which allows for greater control of the aerodynamic forces.

\begin{figure}[htb]
    \centering
    \includegraphics[width=\linewidth]{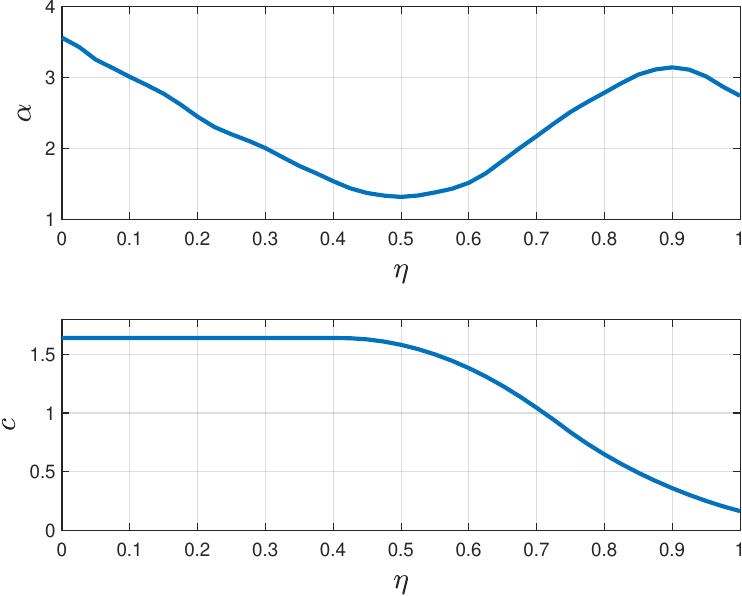}
    \caption{Local cruise angle of attack, $\alpha$, and chord length, $c$, distributions as a function of the relative spanwise position $\eta$, of wing optimized without the strut and optimizing the local twist and chord length, as well as the internal structure.}
    \label{fig:nostrut_twist_chord_distrib}
\end{figure}

The local cruise angle of attack and chord length distributions along the span are shown in \cref{fig:nostrut_twist_chord_distrib}.  The distributions show very moderate twist angles compared to the ones seen in the previous cases, which might be linked to this case resulting in the lightest wing, hence necessitating less lift. From the root to approximately the half span position, the chord length is at its maximum admissible value, and the twist is decreasing, similarly to what was observed in \cref{fig:strut_twist_chord_hist}, when considering the strut. Beyond that position, the chord length decreases and reaches the minimal value at the tip. In most of the region with decreasing chord length, the local twist is increasing again towards a maximum near the spanwise position $\eta=0.9$, beyond which it is again decreasing. This non\ADD{-}trivial twist distribution is linked to the chord length distribution, where the local shape is controlled by the twist. At the half span, the chord length begins to be reduced, and it could therefore be necessary to increase the local lift, by increasing the local twist. Such a configuration, with a higher twist but reduced chord length requires less material in the skin and hence leads to lighter wings. This might also indicate an advantage of airfoils with shorter chord lengths but higher twist angles at these positions. \RM{ Furthermore, the increased twist angles could also increase the bending stiffness in parts of the wing, where the chord length is small. }Near the very tip of the wing ($\eta=0.9$), the twist is again decreased, as the lift at the tip should be minimal.

\begin{figure}[htb]
    \centering
    \includegraphics[width=\linewidth]{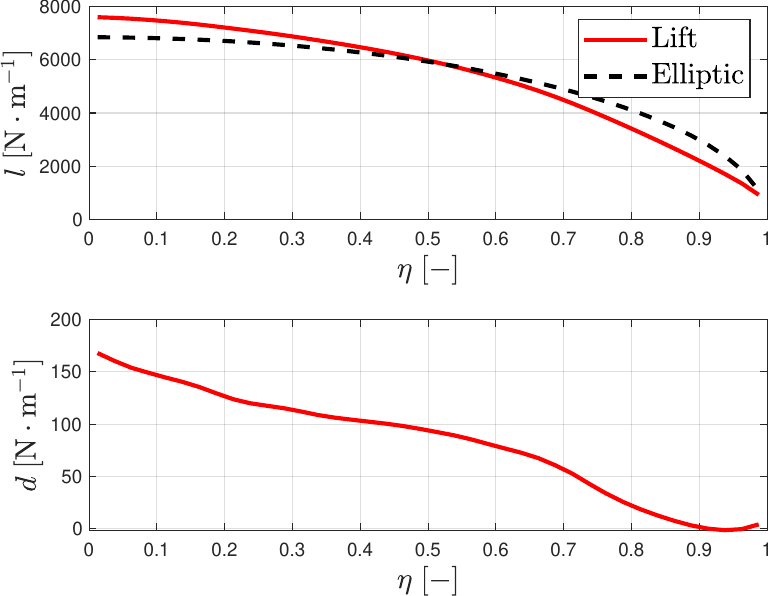}
    \caption{Lift, $l$, and drag, $d$ distributions of the wing optimized without the strut, considering the local twist and chord length, as well as the internal structure.}
    \label{fig:nostrut_twist_chord_lift}
\end{figure}

The lift and drag distributions are shown in \cref{fig:nostrut_twist_chord_lift}, where it is seen that the lift distribution is much closer to an elliptic distribution than the ones of the previous wings, which indicates that the optimization leads to a design that, from an aerodynamic perspective, is close to a structurally unconstrained optimal lift to drag ratio. The lift curve is still skewed towards the root of the wing, compared to an elliptic profile, which indicates that a higher portion of the load is wanted close to the root of the wing, to avoid high bending moments. It is noted that the oscillation in the twist angle, combined with the chord length distribution, seen in \cref{fig:nostrut_twist_chord_distrib}, does not seem to let the lift distribution oscillate, which points towards the chord length and twist having a combined effect to provide the necessary lift while maintaining a low drag and giving loads, that do not infringe on the compliance constraints.

\begin{figure*}[htb]
    \centering
    \begin{subfigure}{0.45\linewidth}
    \includegraphics[width=\linewidth]{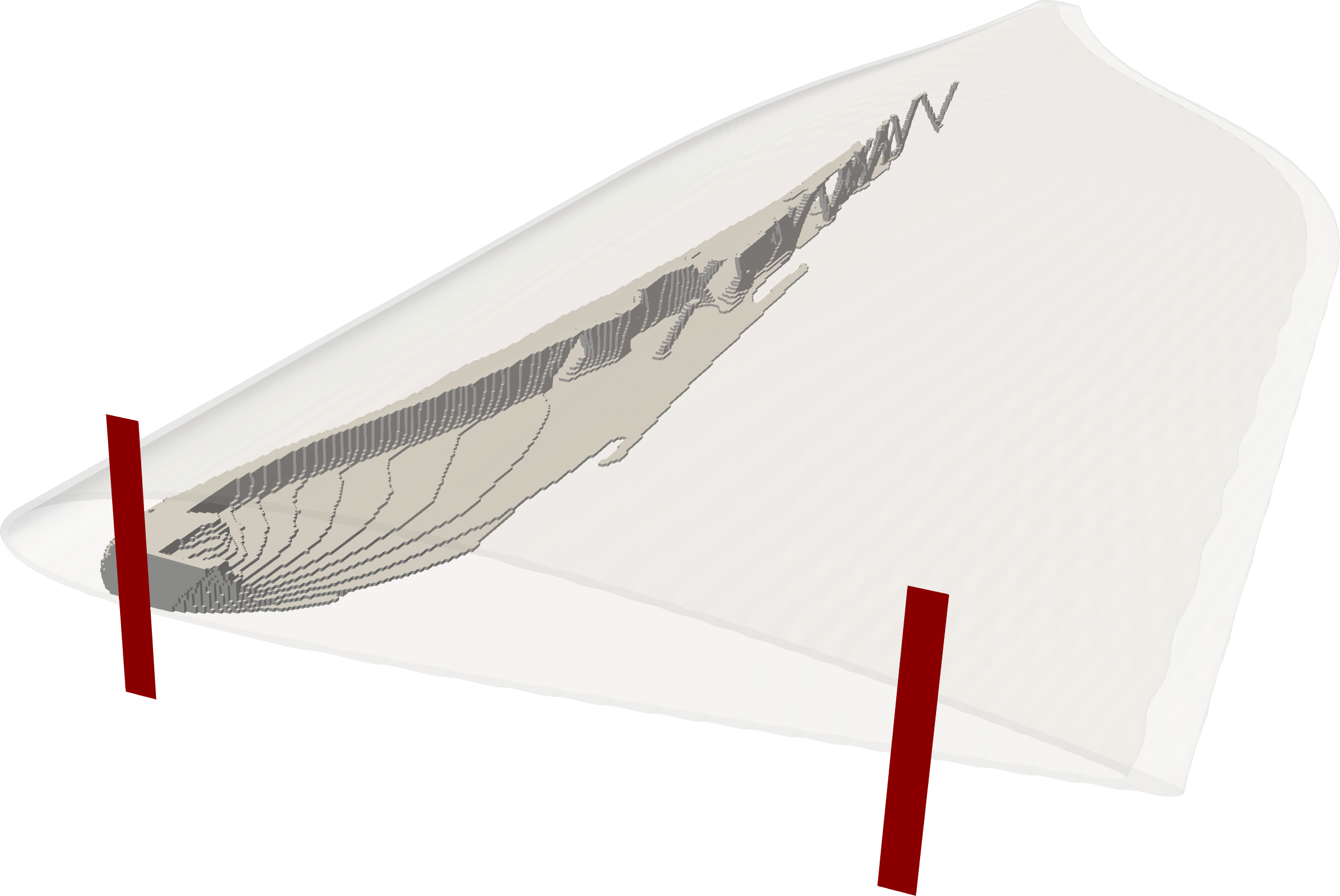}
    \caption{ }
    \end{subfigure}~
    \begin{subfigure}{0.45\linewidth}
    \includegraphics[width=\linewidth]{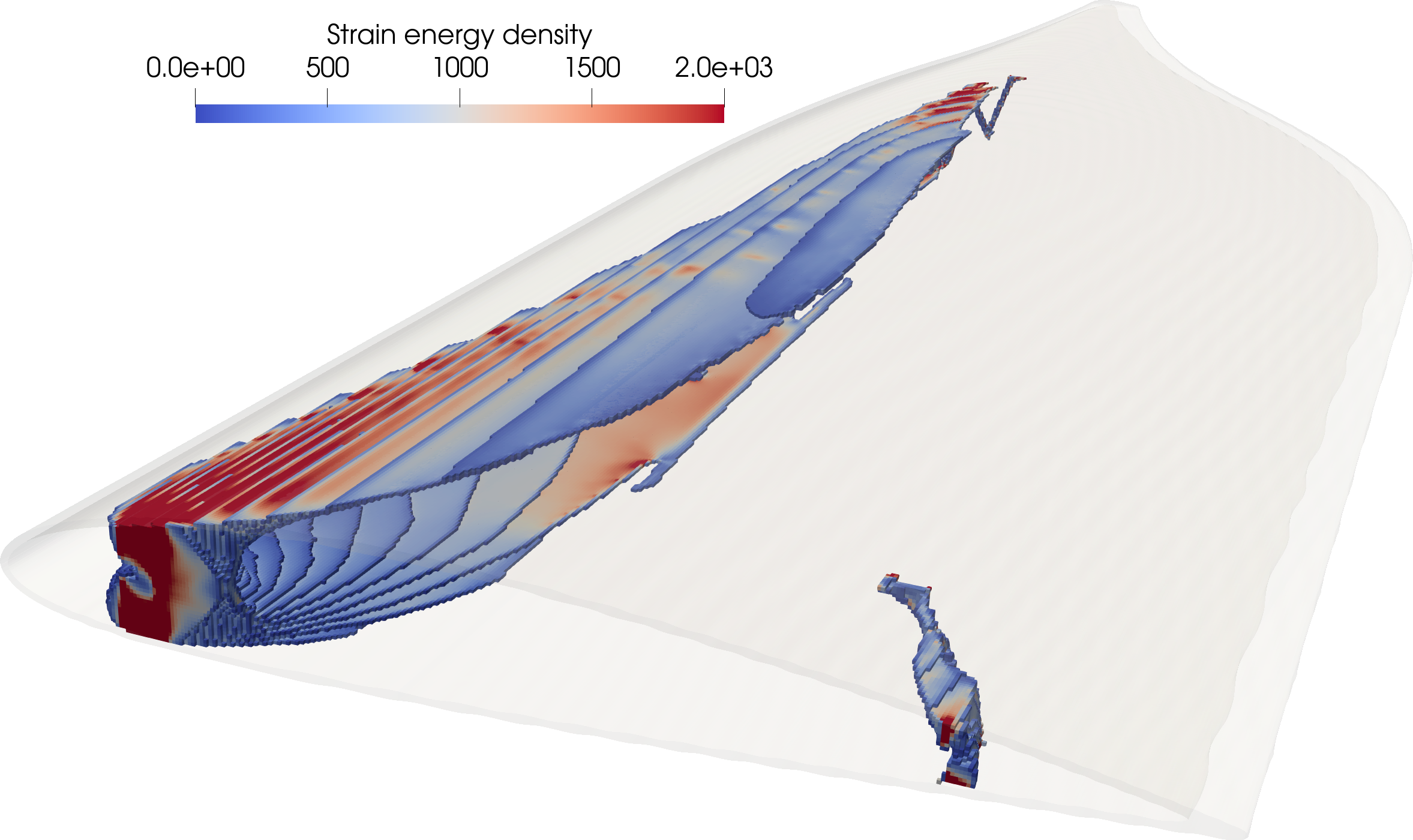}
    \caption{ }
    \end{subfigure}
     \caption{Structure of the wing optimized without the strut and varying twist, chord length, and internal structure. The figures show the fixed thickness skin in a transparent representation and (a) the sliced internal structure (i.e. only the lower part) and (b) the internal structure colored by strain energy density (adjusted for aluminum material properties) in $\mathrm{J\cdot m^{-3}}$. The regions with the clamped boundary conditions are depicted in red in (a).}
    \label{fig:nostrut_twist_chord_struct}
\end{figure*}

The structure of the wing optimized without the strut and considering both the local twist and chord length for the shape problem is shown in \cref{fig:nostrut_twist_chord_struct}. The obtained structure is similar to the one obtained for the same case, optimized without the strut, seen in \cref{fig:nostrut_twist_struct}, with a main I-beam shaped structure connected to the clamped support near the leading edge and a secondary structure at the support near the trailing edge. The connection between the upper and lower skin extends further towards the tip of the wing as in the previous case, without the strut and only optimizing the shape through the local twist. Towards the tip, the connection between the upper and lower skin goes from a plate-like shape to a truss-like structure, which might indicate that the mesh and imposed length scale are too coarse to provide for a thinner plate. A reason for the horizontal stiffener being longer in this case, than when not including the chord length in the optimization could be that the skin patch at the tip of the wing, which acts as a vertical stiffener, is smaller with small chord lengths.

\ADD{The displacements of the wing are corrected for the aluminum Young's modulus and monitored in the final design. The maximum displacement magnitudes in the skin are found as $\max\left(\left|u_\mathit{cruise}\right|\right)=27.8\;\mathrm{mm}$ and $\max\left(\left|u_\mathit{takeoff}\right|\right)=52.9\;\mathrm{mm}$, under cruise and takeoff conditions respectively, which is below $1.1\%$ of the wing span, hence the assumption of small displacements in the optimized wing is still valid.}

\subsection{Comparison of results}
In addition to increased design freedom on the aerodynamical performance of the wing, letting the chord length vary also allows for lighter wings. This is due to the ability to reduce the surface area of wing skin and thereby reducing weight - hence giving further room for improvement of the drag of the wing, as the required lift is now lesser. The three considered types of design variables are interwoven and all three affect, directly or indirectly the objective and all three constraints.

A comparison of the aerodynamic performances at cruise and takeoff, as well as the weight of the initial and optimized wings, is given in \cref{tab:compare}. Here, it is highlighted that a considerable reduction in structural weight of the wing can be achieved - $40\%$ and $36\%$ in the case with and without the strut, respectively when the chord length also is a design variable. The reduction in weight means that less lift is required, which means that the drag is reduced by $31\%$ and $32\%$, respectively. It should be noted that these reductions shouldn't solely be attributed to weight reduction, as the optimization is an integrated process.

The volume fractions of material (on the nominal design) inside the wing are also provided in \cref{tab:compare}. The volume fractions represent the volume of material inside the domain enclosed by the skin. The volume fractions of the two cases considering the chord length as a design parameter are lower than the one of their respective corresponding examples, which do not consider the chord length for the optimization. This reduction in volume fraction coincides with a smaller volume enclosed by the wing skin for the cases where the local chord length is altered, which underlines that the added design freedom must provide large benefits, as both smaller and less dense wings are obtained in these cases.

\begin{table*}[htb]
\centering
\caption{Comparison of the optimized and initial  wings}
\label{tab:compare}
\begin{tabular}{l|rrrrrr}
Optimization & Cruise drag  &Cruise lift & Takeoff drag & Takeoff lift & Structure weight & Volume fraction \\ \hline
Initial  & $759\;[\mathrm{N}]$ & $18.7\;[\mathrm{kN}]$ & $1510\;[\mathrm{N}]$ & $26.3\;[\mathrm{kN}]$ & $13.4\;[\mathrm{kN}]$ & $50\%$\\
With strut, twist & $369\;[\mathrm{N}]$ & $9.27\;[\mathrm{kN}]$ & $785\;[\mathrm{N}]$ & $17.1\;[\mathrm{kN}]$ & $3.27\;[\mathrm{kN}]$ & $10.5\%$\\
With strut, chord, twist & $281\;[\mathrm{N}]$ & $7.96\;[\mathrm{kN}]$ & $630\;[\mathrm{N}]$ & $14.7\;[\mathrm{kN}]$ & $1.96\;[\mathrm{kN}]$ & $8.6\%$\\
Without strut, twist & $166\;[\mathrm{N}]$ & $7.74\;[\mathrm{kN}]$ & $536\;[\mathrm{N}]$ & $15.5\;[\mathrm{kN}]$ & $1.74\;[\mathrm{kN}]$ & $5.2\%$\\
Without strut, chord, twist & $113\;[\mathrm{N}]$ & $7.11\;[\mathrm{kN}]$ & $410\;[\mathrm{N}]$ & $13.6\;[\mathrm{kN}]$ & $1.11\;[\mathrm{kN}]$ & $4.8\%$\\
\end{tabular}
\end{table*}

Comparing the internal structures obtained when considering the strut and when not, it is noted that the structures are very different. This comparison should be made carefully, as the compliance constraints are not identical. However, one can note that when the strut is included in the model, the structure branches into- and out of the connection point. In both distinct cases, with and without a strut, the structures seem to be qualitatively similar, as there is some structure connecting the upper and lower skin going through the wing and transferring the loads to the strut and the support in the structures where the strut is considered, seen in \cref{fig:strut_twist_struct,fig:strut_twist_chord_struct}. In the two cases without the strut, seen in \cref{fig:nostrut_twist_struct,fig:nostrut_twist_chord_struct}, the structure primarily only consists of an I-beam like structure connecting the upper and lower skin.

In all cases the strain energy density appears to be relatively evenly distributed throughout the wing, \ADD{except for} the regions near the supports. This underlines that the used material is indeed necessary to keep the compliances at the specified level.
\section{Conclusion}
\label{sec:conc}
The optimization of the external shape of wings as well as the topology of their internal structure has successfully been carried out. One-way coupled physics were applied, where the change in aerodynamic performance due to the wing deflection was neglected.

The state evaluation is carried out in a segregated manner: first, the panel method is used to solve the aerodynamic response of the wing, then the aerodynamic forces are projected to the unfitted mesh used for the structural response evaluation and optimization, finally, the structural response of the wing is evaluated using linear elastic finite elements.

Two different wing cases were considered one with a wing strut, inspired from small, single-engine airplanes, and another one without the strut. For each case, two optimizations were carried out: one where the shape only is optimized through the local twist, and the other, where both the local twist and chord length are optimized for the wing shape. The increased design freedom, when also considering the local chord length in the optimization, leads to a reduction in drag by up to $32\%$.

The optimization can have some trouble finding a feasible design, as the objective and three constraint functions are contradicting to a very high degree. For instance, if the imposed maximum compliances are too small, the optimization satisfies them, but at the cost of a too-low lift, such that the weight-lift constraint ($g_1$) is violated. Generally, the optimization process with intertwined objective and constraint functions having opposite motivations for the final design is seen to be difficult. The optimization requires tight asymptote settings and can get locked in infeasible designs under certain conditions.

It should be noted that important dynamic effects, such as flutter are ignored in this study. These effects would also influence the structural design. \ADD{Buckling of the wing skin is also a major concern for the optimized structures. Future work could concern the implementation of buckling constraints in the presented optimization approach.}

Structural constraints based on compliance are easily implemented and computationally efficient. However, stress constraints can directly be linked to the material properties and safety factors, hence rendering the optimization a more integrated part of the design process. Having a set of few discrete load cases  (from different angles of attack) does not seem to be sufficient to avoid regions with low bending moments in all cases, as seen in the cases, where the strut is included, where regions with no or very little material appear between the strut and root of the wing. These regions with no material could also be avoided with finer meshes, which would allow for thinner wing skins, with a realistic thickness. Furthermore, the skin could then be modeled with the same material properties as the internal structure. Additionally, a finer discretization level would also allow for smaller features inside the wing (if the length scale is adapted accordingly). Finally, if a higher fidelity description of the aerodynamics or more advanced aerodynamic modeling was used, where viscous forces are included, the airfoil profile at each section could also be included in the optimization.

\appendix
\section{Sensitivity analysis}
\label{app:sens}
Differentiating the augmented Lagrangian function $\mathcal{F}$ from \cref{eq:auglag} with respect to the shape- and structural design variables, $d$ and $\gamma$, respectively, and applying the chain rule:
\begin{equation}
\begin{split}
   \frac{d\mathcal{F}}{d\gamma} = \frac{\partial f}{\partial\gamma}&+\frac{\partial f}{\partial\mathbf{u}}\frac{d\mathbf{u}}{d\gamma}+\bm{\lambda}_s^\intercal\left(\frac{\partial\mathbf{r}_s}{\partial\gamma}+\frac{\partial\mathbf{r}_s}{\partial\mathbf{u}}\frac{d\mathbf{u}}{d\gamma}\right)\\
    \frac{d\mathcal{F}}{dd^s_i} = \frac{\partial f}{\partial d^s_i}&+\frac{\partial f}{\partial\mathbf{u}}\frac{d\mathbf{u}}{dd^s_i}+\frac{\partial f}{\partial\bm{\mu}}\frac{d\bm{\mu}}{dd^s_i}\\&+\bm{\lambda}_s^\intercal\left(\frac{\partial\mathbf{r}_s}{\partial d^s_i}+\frac{\partial\mathbf{r}_s}{\partial\mathbf{u}}\frac{d\mathbf{u}}{dd^s_i}+\frac{\partial\mathbf{r}_s}{\partial\bm{\mu}}\frac{d\bm{\mu}}{dd^s_i}\right)\\&+\bm{\lambda}_a^\intercal\left(\frac{\partial\mathbf{r}_a}{\partial d^s_i}+\frac{\partial\mathbf{r}_a}{\partial\bm{\mu}}\frac{d\bm{\mu}}{dd^s_i}\right)
\end{split}
\end{equation}
The above expressions can be rewritten by taking advantage of the arbitrary choice of Lagrangian multipliers, which allows setting certain terms to zero:
\begin{equation}
\begin{split}
    \frac{d\mathcal{F}}{d\gamma} = \frac{\partial f}{\partial\gamma}&+\bm{\lambda}_s^\intercal\frac{\partial\mathbf{r}_s}{\partial\gamma}+\underbrace{\left(\frac{\partial f}{\partial\mathbf{u}}+\bm{\lambda}_s^\intercal\frac{\partial\mathbf{r}_s}{\partial\mathbf{u}}\right)}_{=0}\frac{d\mathbf{u}}{d\gamma}\\
    \frac{d\mathcal{F}}{dd^s_i} = \frac{\partial f}{\partial d^s_i}&+\bm{\lambda}_a^\intercal\frac{\partial \mathbf{r}_a}{\partial d^s_i}+\bm{\lambda}_s^\intercal\frac{\partial\mathbf{r}_s}{\partial d^s_i}\\&+\underbrace{\left(\frac{\partial f}{\partial \bm{\mu}}+\bm{\lambda}_a^\intercal\frac{\partial\mathbf{r}_a}{\partial\bm{\mu}}+\bm{\lambda}_s^\intercal\frac{\partial\mathbf{r}_s}{\partial\bm{\mu}}\right)}_{=0}\frac{d\bm{\mu}}{dd^s_i}\\&+\underbrace{\left(\frac{\partial f}{\partial\mathbf{u}}+\bm{\lambda}_s^\intercal\frac{\partial\mathbf{r}_s}{\partial\mathbf{u}}\right)}_{=0}\frac{d\mathbf{u}}{dd^s_i}
\end{split}
\end{equation}
Hence, in the adjoint system of equations, to retrieve the Lagrangian multipliers $\bm{\lambda}_a$ and $\bm{\lambda}_s$, the one-way coupling is reversed, such that the structural adjoint affects the aerodynamic adjoint - but not the other way around.

The chain rule for the structural design field is given as:
\begin{align}
    \frac{d\rho_e}{d\gamma} &= \underline{\xi}\frac{\partial\gamma_e}{\partial\tilde{\gamma}}\frac{d\tilde{\gamma}}{d\gamma} & \frac{d\rho_e}{dd^s_i} &=\left( \frac{d\overline{\xi}}{d\tilde{\xi}}+\frac{d\underline{\xi}}{d\tilde{\xi}}\left(\gamma_e-1\right)\right)\frac{d\tilde{\xi}}{d\xi}\frac{d\xi}{dd^s_i}
\end{align}
The chain rule term of the smooth Heaviside projection is given as:
\begin{equation}
    \frac{d\underline{\xi}}{d\tilde{\xi}}=\beta\frac{1-\tanh^2(\beta(\tilde{\xi}-1))}{\tanh(\beta\eta)+\tanh(\beta(1-\eta))}
\end{equation}

\section*{Acknowledgements}
We acknowledge the financial contributions from the Villum Foundation through the Villum Investigator project InnoTop.

\section*{Conflict of interest}
On behalf of all authors, the corresponding author states that there is no conflict of interest. 

\section*{Replication of results}
The methods used to produce the results of this paper have been outlined in \cref{sec:param,sec:phys,sec:opt}. Additional implementation details can be provided upon request.
\bibliographystyle{spbasic}
\bibliography{bib.bib}   
\end{document}